\documentclass[english,preprint]{aastex}
\usepackage{mathptmx}

\usepackage[T1]{fontenc}
\usepackage[latin9]{inputenc}
\usepackage{color}
\usepackage{amssymb}
\usepackage{graphicx}
\usepackage{esint}

\makeatletter

\usepackage{graphicx}

\def\aap{{Astronomy and Astrophys.}}	

\def\araa{{ARA\&A}}

\def\mnras{{MNRAS}}

\def\apj{{\it Astrophys. J.\ }}
\def\apjs{{\apj\ \it Suppl.\ }}		
\def\apjl{{\apj\ \it Lett.\ }}

\def\mnras{{\it Mon. Not. R. Astron. Soc.\ }}
	
\def\ssr{{\it Space Sci. Rev.\ }}

\def\eg{{ e.g.,\ }}

\def\ie{{i.e.,\ }}

\def\ln{{\rm ln}}
\usepackage{hyperref}

\makeatother

\usepackage{babel}
\begin{document}

\title{Analytic Solution for Self-regulated Collective Escape of Cosmic
Rays from their Acceleration Sites}

\author{M.A. Malkov and P.H. Diamond%
\footnote{Also at WCI Center for Fusion Theory, NFRI, Korea%
}}

\affil{CASS and Department of Physics, University of California, San Diego,
La Jolla, CA 92093}

\email{mmalkov@ucsd.edu, pdiamond@ucsd.edu}

\author{R.Z. Sagdeev}

\affil{University of Maryland, College Park, Maryland 20742-3280, USA}

\author{F.A. Aharonian}

\affil{Dublin Institute for Advanced Studies, 31 Fitzwilliam Place, Dublin
2, Ireland }

\affil{Max-Planck-Institut f\"ur Kernphysik, PO Box 103980, 69029 Heidelberg,
Germany}

\author{I.V. Moskalenko}

\affil{Hansen Experimental Physics Laboratory, Stanford University, Stanford,
CA 94305}
\begin{abstract}
Supernova remnants (SNRs), as the major contributors to the galactic
cosmic rays (CR), are believed to maintain an average CR spectrum
by diffusive shock acceleration (DSA) regardless of the way they release
CRs into the interstellar medium (ISM). However, the interaction of
the CRs with \emph{nearby} gas clouds crucially depends on the release
mechanism. We call into question two aspects of a popular paradigm
of the CR injection into the ISM, according to which they \emph{passively}
and \emph{isotropically} diffuse in the prescribed magnetic fluctuations
as test particles. First, we treat the escaping CR and the Alfven
waves excited by them on an equal footing. Second, we adopt field
aligned CR escape outside the source, where the waves become weak.
An exact analytic self-similar solution for a CR ``cloud\textquotedbl{}
released by a dimmed accelerator strongly deviates from the test-particle
result. The normalized CR partial pressure may be approximated as
$\mathcal{P}\left(p,z,t\right)=2\left[\left|z\right|^{5/3}+z_{{\rm dif}}^{5/3}\left(p,t\right)\right]^{-3/5}\exp\left[-z^{2}/4D_{{\rm ISM}}\left(p\right)t\right]$,
where $p$ is the momentum of CR particle, and $z$ is directed along
the field. The core of the cloud expands as $z_{{\rm dif}}\propto\sqrt{D_{{\rm NL}}\left(p\right)t}$
and decays in time as $\mathcal{P}\propto2z_{{\rm dif}}^{-1}\left(t\right)$.
The diffusion coefficient $D_{{\rm NL}}$ is strongly suppressed compared
to its background ISM value $D_{{\rm ISM}}$: $D_{{\rm NL}}\sim D_{{\rm ISM}}\exp\left(-\Pi\right)\ll D_{{\rm ISM}}$
for sufficiently high field-line-integrated CR partial pressure, $\Pi$.
When $\Pi\gg1$, the CRs drive Alfv\'en waves efficiently enough to
build a \emph{transport barrier} ($\mathcal{P}\approx2/\left|z\right|$
-``pedestal'') that strongly reduces the leakage. The solution has
a spectral break at $p=p_{{\rm br}}$, where $p_{{\rm br}}$ satisfies
the following equation $D_{{\rm NL}}\left(p_{{\rm br}}\right)\simeq z^{2}/t$. 
\end{abstract}

\section{Introduction}

The generation of cosmic rays (CR) in supernova remnant (SNR) shocks
by the diffusive shock acceleration (DSA) mechanism \citep[e.g., ][]{DruryNetw01}
is understood reasonably well up to the point of their escape into
SNR surroundings. But making this mechanism\textcolor{blue}{{} }responsible
for the most of galactic CRs requires understanding of all stages
of the CR production including their escape from the accelerators.
In fact, best markers for ``CR-proton factories'' are nearby molecular
clouds (MC) illuminated by protons leaking from SNRs. CRs will be
visible in gamma rays generated by collisions with protons in the
cloud \citep{AharDV94,AharAt96,AharW28HESS08,GabiciAharEsc09,TavaniIC443_10,Abdo10W44full,AbdoW28_10,AgileW44_11,EllisonBykEscape11,Torres11}.
Whether this gamma radiation is detectable with the current instruments,
depends on the CR leakage rate from the source. The recent surge in
measurements of gamma-bright SNR suggests that the sensitivity threshold
have already been surpassed for at least several galactic SNRs and
it is becoming increasingly timely to improve our understanding of
the CR leakage from these objects. 

Without such improvement, it is also difficult to resolve the ongoing
debates about the primary origin of gamma-emission from some of the
gamma-active remnants in complicated environs, \eg RX J 1713 \citep[e.g.,][]{FunkRapport12,Inoue12,FermiSciW4413}.
In arguing for hadronic or leptonic origin, one needs to know exactly
how far the CRs are spread from the source at a given time and with
what spectrum. Indeed, strong self-confinement of accelerated CR may
keep their flux through a remote MC below the instrument threshold,
primarily (and counterintuitively) for powerful accelerators. Conversely,
the self-confinement will enhance illumination of nearby MCs, thus
enhancing the odds of detecting the hadronic contribution to the emission.
Apart from the distance to the target MC, equally important is its
magnetic connectivity with the CR source. Overall, the predictions
for the emissivity of MCs near strong CR sources can differ from the
test-particle results by an order of magnitude or more. 

To better understand the physics of CR confinement to SNRs, we consider
the CR escape separately from their acceleration which is assumed
to have faded because of the SNR age. Specifically, we will formulate
the problem as a diffusion of a CR cloud (CRC) released from an accelerator
into the ISM and propagating through a 'gas' of self-excited Alfv\'en
waves. At the scales larger than the initial size of the cloud, the
solution, after an adjustment to the local environments, will become
self-similar to depend only on the background diffusivity $D_{{\rm ISM}}$
and the integrated CRC energy (cf. Sedov-Taylor solution for the point
explosion). 

The idea that the CRC confinement should be thought of as a \emph{self-confinement}
is not new \citep[e.g.,][]{Wentzel74,Acht81a}. However, the analytic
solution for CR propagation uniformly valid in both the nearby and
far zones of the initial CRC seems to be unknown. This paper is aimed
at presenting and analyzing such solution. With some reservations,
it may also be applied to a late stage of CR acceleration in a SNR,
when the accelerated CRs become disconnected from the shock and diffuse
outwards. Their pressure, however, is still sufficient to drive strong
Alfv\'en waves that limit the CR escape. We begin with a brief discussion
of the problems with the current escape models, thus motivating the
new one.

\subsection{The need for a new CR escape model}

Traditionally, the transport of CR is treated differently inside and
outside a DSA accelerator. Outside, CRs are thought to escape as \emph{test
particles} with a diffusion coefficient inferable from observations.
This picture cannot be correct near the accelerator because the CR
transport must be in a \emph{self-confinement }regime\emph{,} in which
the CR streaming instability diminishes their diffusion coefficient
by orders of magnitude. This CR anisotropy instability, and particularly
its nonresonant extensions, macroscopically driven by the CR current
and the CR pressure gradient, have a potential to generate magnetic
field fluctuations well in excess of the ambient magnetic field, $\delta B\gtrsim B_{0}$
\citep[e.g.,][]{Bell04,DruryFal86,BykovBell09,MDS10PPCF}. At the
very least, such strong fluctuations should justify the standard DSA
assumption about the Bohm diffusion regime with the mean free path,
(mfp) of the order of the particle gyroradius, $r_{{\rm g}}$, achieved
at $\delta B\sim B_{0}$. Naturally, the transport is isotropic in
this regime. The ISM background turbulence, on the other hand, is
much weaker $\delta B^{2}/B_{0}^{2}\lesssim10^{-5}$ at the CR relevant
length scales, thus resulting in the CR mfp $\gtrsim10^{5}r_{{\rm g}}$
(for GeV particles, $r_{{\rm g}}\gtrsim10^{12}$cm). It is important
to emphasize that under these circumstances the cross-field diffusion
coefficient, $\kappa_{\perp}$ is suppressed by a large factor compared
to the diffusion along the field line, $\kappa_{\parallel}$, i.e.
$\kappa_{\perp}/\kappa_{\parallel}\sim\left(\delta B/B_{0}\right)^{4}$
(see, e.g., \citealt{Drury83}). 

It follows then that there is a problem of describing particle transport
between the self-confinement (accelerator vicinity) and the test-particle
(far from accelerator) transport regimes. To circumvent this problem,
the acceleration process has been treated separately from the particle
escape using one of the two devices: the upper cut-off momentum and
the free-escape boundary (FEB) \citep[see e.g., ][for recent discussions]{RevilleEsc09,DruryEscape11}.
As the names suggest, accelerated particles escape instantly upon
reaching a prescribed boundary either in momentum or in configuration
space. Their escape is assumed to have no effect on the acceleration
other than through the modification of the shock structure \citep{Mosk07}.
In a simplified visualization of the DSA as a 'box' process, for example
\citep{DruryBox99}, the upper cutoff and the FEB are two sides of
the box in the particle phase space. There were efforts to include
self-generated waves into the description of particle acceleration
and escape from strongly modified CR shocks \citep{mdj02,MD06}, or
in numerical treatments \citep{Galinsky07,GalShev11,Fujita11}. In
most approaches, however, a sudden jump in the CR diffusion coefficient
in momentum space is introduced to set an upper cutoff, \citep[e.g., ][]{PtusZir05,YanLazarianEscape12}.
Similar jump in coordinate space would result in a FEB.

As long as the CR transport outside the accelerator is treated in
the test particle approximation, no smooth transition between the
CR acceleration and their escape is provided, regardless of the scenario
for the latter. But, the diffusion coefficient rises by roughly five
orders of magnitude in a transition zone that should be correspondingly
large, unlike an infinitely thin FEB. It is this zone where the CR
escape flux and confinement time are set by self-generated waves,
thus rendering the FEB a rather implausible concept. For many SNRs
it is then not yet possible to conclude whether the gamma-emission
is leptonic or hadronic, should such conclusion depend on the CR escape
and on the subsequent illumination of adjacent clouds. In a broader
sense, there is a missing link in galactic CR generation between the
CR acceleration (under a very strong self-generated wave-particle
scattering) and their subsequent propagation (in a very weak interstellar
turbulence). The goal of this paper is to establish such link.

Our treatment below is applicable to the following two situations.
In one situation, a shock accelerates particles continuously but some
of them reach far enough or diffuse across the local field to become
disconnected from the shock front and have thus chances to escape.
While doing so, they drive their own waves at a gyroradius scale,
whose amplitudes gradually decrease outwards and so does the particle
density. The second situation is a clear-cut case for this paper that
deals with a CRC released into the ISM with no ongoing acceleration
inside the CRC. Both situations correspond to a mesoscale transport
regime, intermediate between the Bohm diffusion inside the accelerator
and a global, quasi-isotropic CR diffusion in the Galaxy. The latter
process is, in turn, supported by a random large scale component of
the galactic magnetic field at the scale of tens of parsecs superposed
on the regular, spiral arm aligned toroidal field \citep{BrunettiProp00,MoskStrong07ARNPS,BlasiChemComp12}.
In the mesoscale transport regime, however, these components are considered
as an ambient field with a scale much larger than the CR gyroradius
and even larger than the CRC scale height. Before we describe in the
next section the physical setting of the problem, it is worthwhile
to emphasize some of the qualitative results.

First of all, for sufficiently high initial CR pressure (higher than
magnetic pressure) and low ambient turbulence level, $\delta B\ll B_{0}$,
the spreading of the CRC strongly deviates from the oft-used test-particle
solution. Instead of being controlled by only one (diffusive) scale
$l_{{\rm D}}\sim\sqrt{D_{{\rm ISM}}t}$, the structure of the expanding
CRC is more complicated. Namely, it comprises three zones, of which
only the outermost has the above test-particle scaling but, a significantly
\emph{lower }CR pressure than it would maintain by diffusing through
the ISM with the diffusivity $D_{{\rm ISM}}$. Conversely, the CR
pressure in the innermost part remains strongly enhanced. These two
zones are connected by a self-similar, $1/z$-part of the CR pressure
profile.

\section{CR Escape Model}

After the Sedov-Taylor (ST) expansion stage in which the shock radius
increases as $R_{{\rm s}}\propto t^{2/5}$ and particularly in and
after the so-called pressure driven snowplow stage with a slower expansion
at or below $R_{{\rm s}}\propto t^{0.3}$ \citep[e.g.,][]{Bisnovatyi95,TrueloveMcKee99},
the diffusive propagation of CR away from the shock becomes more important
than their bulk expansion driven by the overpressured SNR interior.
Regardless of the escape mechanism, one of the most important parameters
is the number of spatial dimensions involved (see, e.g., \citealp{DruryEscape11}
for a recent discussion). Indeed, the 3D random walk is non-recurrent,
so only the finite shock radius gives particle a chance to return
to it. However, 2- and 1-D random walks are recurrent. Therefore,
the first important assumption to make is, indeed, about the process
dimensionality which immediately translates into the choice of magnetic
field configuration. Note that since particles recede from the shock
as $\bar{R}\propto t^{1/2}$, they escape already at the Sedov-Taylor
stage. 

Contrary to the recent analytic \citep{DruryEscape11,OhiraProp11,YanLazarianEscape12}
and numerical (with self-driven Alfv\'en waves \citealp{Fujita11})
treatments of the spherically symmetric particle escape, we consider
an escape through the local flux tube. This choice is suitable for
a SNR expansion in an ISM with a distinct large scale magnetic field
direction that does not change very strongly on the SNR scale. As
we stated in the Introduction, our goal is to fill the gap between
the weak scattering propagation far away from the remnant and the
Bohm diffusion regime near the shock, with $\kappa=\kappa_{{\rm B}}$.
The perpendicular diffusion in the far zone is of the order of $\kappa_{\perp}\sim\left(\delta B/B_{0}\right)^{2}\kappa_{{\rm B}}\ll\kappa_{{\rm B}}\ll\kappa_{{\rm \parallel}}\sim\left(\delta B/B_{0}\right)^{-2}\kappa_{{\rm B}}$,
so that we may safely assume that in the unperturbed ISM the diffusive
propagation is one-dimensional, along the field line.

\begin{figure}
\includegraphics[scale=0.6]{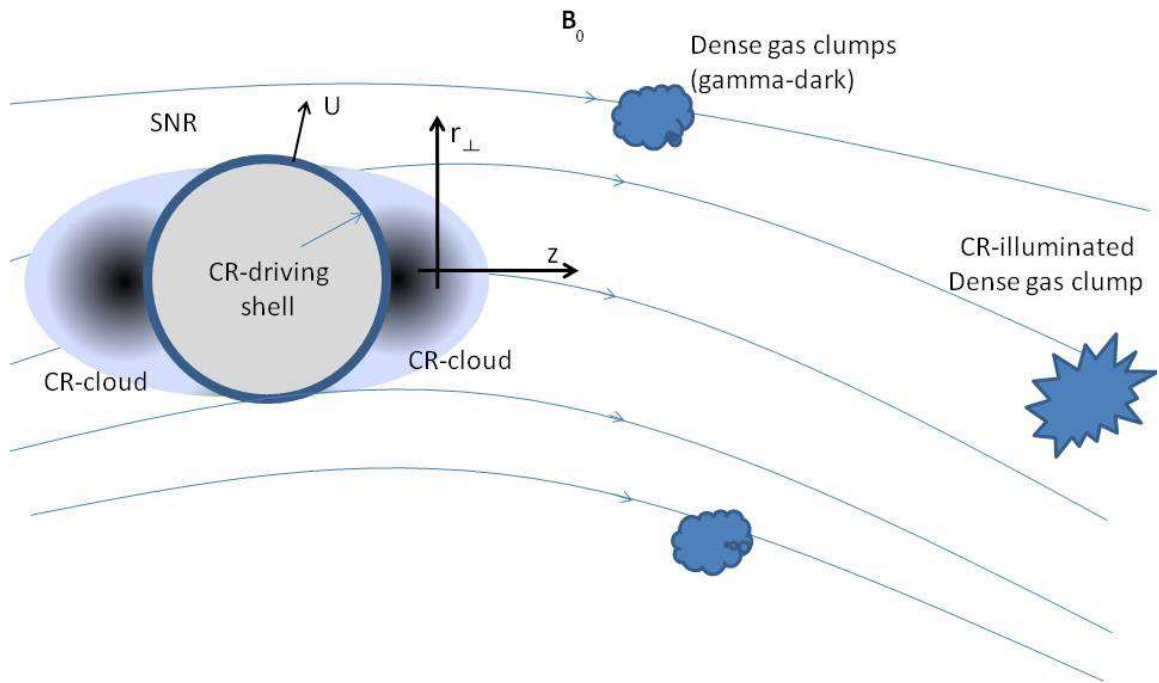}

\caption{CR escape along the magnetic field $\mathbf{B}_{0}$ from the two
polar cusps of SNR with a stalled blast wave.\label{fig:CR-escape-along}}
\end{figure}

Closer to and inside the accelerator (or rather in the region of its
past activity), $\kappa_{\perp}\sim\kappa_{\parallel}\sim\kappa_{{\rm B}}$,
which favors the isotropy assumption. However, we may integrate (average)
the equations for the CR transport and wave generation across the
magnetic flux tube. Then, the problem becomes, again, formally one
dimensional. However, the lateral expansion of the integrated flux
tube may decrease CR pressure inside thus making the CR self-confinement
in the field direction less efficient.To estimate this effect we compare
the 'perpendicular' confinement time, $\tau_{\perp}\sim a_{\perp}/U_{\perp}$
with the 'parallel' confinement time $\tau_{\parallel}\sim a_{\parallel}^{2}/D_{{\rm eff}}$.
Here $a_{\perp,\parallel}$ denote the respective sizes of the CR
cloud, $U_{\perp}$is the expansion velocity across magnetic field,
and $D_{{\rm eff}}$ is the effective CR diffusion coefficient along
the field. We may estimate $U_{\perp}$ from the balance of the CR
pressure (assuming $P_{{\rm CR}}\gg B^{2}/8\pi$) and the ram pressure
of the ambient medium, $U_{\perp}\sim\sqrt{P_{{\rm CR}}/\rho}$. The
effective diffusivity, associated with the half-life of the CR against
losses in the field direction, is shown in Appendix \ref{sec:Half-life-of-CR}
to be $D_{{\rm eff}}\sim\sqrt{D_{{\rm {\rm ISM}}}D_{{\rm NL}}}$,
where $D_{{\rm NL}}$ is the CR diffusivity suppressed by the self-generated
Alfv\'en waves. Requiring $\tau_{\perp}\gg\tau_{\parallel}$ and estimating
$a_{\perp}\sim R_{{\rm SNR}},$ $a_{\parallel}\sim\kappa_{{\rm B}}/U_{{\rm sh}}$,
we convert the inequality $\tau_{\perp}\gg\tau_{\parallel}$ into
the following constraint on the CR acceleration efficiency $P_{{\rm CR}}/\rho U_{{\rm sh}}^{2}<\left(U_{{\rm sh}}R_{{\rm SNR}}/\kappa_{{\rm B}}\right)^{2}\left(D_{{\rm NL}}D_{{\rm ISM}}/\kappa_{{\rm B}}^{2}\right)$.
As both factors in the parentheses on the r.h.s. are larger than unity,
the above requirement is fulfilled. The CRs then indeed escape along
the field line before they inflate the flux tube significantly (cf.
e.g. \citealp{Rosner96,PtuskinNLDIFF08}, where the field aligned
propagation from CR sources has also been adopted). A cartoon example
of the basic configuration is shown in Fig.\ref{fig:CR-escape-along}.
It should be noted that our simplified CR propagation scenario does
not inlude possibly important perpendicular CR transport due to the
CR drifts, magnetic field meandering or turbulence spreading. 

The second important assumption to make is about the spatial arrangement
of the initial CR population. As shown in Fig.\ref{fig:CR-escape-along},
we adhere to the idea that the acceleration is most efficient in the
quasi-parallel shock geometry. It can be advocated on the theoretical
grounds \citep{mv95,VoelkInj03}, by \emph{in situ }observations of
heliospheric shocks,and hybrid simulations \citep[e.g., ][]{Burgess12,SpitkovskyHybr12}.
More importantly, this acceleration preference is supported by SNR
observations, \citep[e.g.][]{Reynoso1006_13}. Therefore, we may specifically
assume that two 'polar cusps' of accelerated particles are left behind
after the acceleration has either faded out or entered its final stage
when particles escape faster then they are replenished by the acceleration.
It is tempting to consider the SN 1006 as a prototype of such geometry,
but the similarity is physically not quite convincing, given the young
age of the latter source. On the other hand, older remnants, such
as W44, do show a bipolar CR escape (\citealp{UchiW44_12}) that can
also be attributed to the field aligned escape.

During earlier, more active stages of acceleration, CR presumably
fill up both the downstream and upstream regions near the shock. Meanwhile,
the contact discontinuity (CD) behind the cloud of accelerated CRs
must have undergone the Rayleigh-Taylor instability with strong magnetic
field enhancement \citep{Gull73}. The CD expansion at late evolution
stages should thus act as a piston on the previously accelerated CRs.
Note that the CR reflecting piston was already employed in numerical
acceleration schemes, \eg \citep{Ber96}, which might, however,overestimate
the maximum CR energy \citep{KirkDendy01}. In the post acceleration
stage, however, given the significant field amplification at the CD,
it is reasonable to assume that CRs are partially coupled to the slowly
expanding flow with the reflecting magnetic piston behind but while
escaping upstream they couple to the ISM and diffuse away.

\subsection{Basic equations, initial and boundary conditions }

Perhaps the most systematic quasilinear derivation of equations for
the coupled evolution of CR and Alfv\'en waves is given by \citet{Skill75a}.
We use them in a simplified one-dimensional (along the ambient field,
$z$-coordinate, Fig.\ref{fig:CR-escape-along}) form equivalent to
that used by, \eg \citet{Bell78} and, with an interesting ``beyond
quasilinear'' interpretation, by \citet{Drury83}

\begin{equation}
\frac{d}{dt}P_{{\rm {\rm CR}}}\left(p\right)=\frac{\partial}{\partial z}\frac{\kappa_{{\rm B}}}{I}\frac{\partial P_{{\rm CR}}}{\partial z}\label{eq:dPdt}
\end{equation}

\begin{equation}
\frac{d}{dt}I=-C_{{\rm A}}\frac{\partial P_{{\rm CR}}}{\partial z}-\Gamma I.\label{eq:dIdt}
\end{equation}
Here $C_{{\rm A}}$ is the Alfv\'en velocity, and the time derivative
is taken along the characteristics of unstable Alfv\'en waves, forward
propagating with respect to the flow of speed $U$:

\begin{equation}
\frac{d}{dt}=\frac{\partial}{\partial t}+\left(U+C_{{\rm A}}\right)\frac{\partial}{\partial z}\label{eq:charct}
\end{equation}
Eq.(\ref{eq:dPdt}) above is essentially a well-known convection-diffusion
equation, written for the dimensionless CR partial pressure $P_{{\rm CR}}$
instead of their distribution function $f\left(p,t\right)$. We have
normalized it to the magnetic energy density $\rho C_{{\rm A}}^{2}/2$:

\begin{equation}
P_{{\rm CR}}=\frac{4\pi}{3}\frac{2}{\rho C_{{\rm A}}^{2}}vp^{4}f,\label{eq:PcrDef}
\end{equation}
where $v$ and $p$ are the CR speed and momentum, and $\rho$- the
plasma density. The total CR pressure is normalized to $d\ln p$,
similarly to the wave energy density $I$: 

\[
\frac{\left\langle \delta B^{2}\right\rangle }{8\pi}=\frac{B_{0}^{2}}{8\pi}\int I\left(k\right)d\ln k=\frac{B_{0}^{2}}{8\pi}\int I\left(p\right)d\ln p
\]

Eq.(\ref{eq:dIdt}) is a wave kinetic equation in which the energy
transferred to the waves equals to the difference between the total
work done by the particles, $\left(U+C_{{\rm A}}\right)\nabla P_{{\rm CR}}$,
and the work done on the fluid, $U\nabla P_{{\rm CR}}$ \citep{Drury83}.
This interpretation of the wave generation indicates that it operates
in a maximum efficiency regime. A formal quasilinear derivation of
this equation assumes that the particle momentum $p$ is related to
the wave number $k$ by the 'sharpened' resonance condition $kp=eB_{0}/c$
instead of the conventional cyclotron resonance condition $kp_{\|}=eB_{0}/c$
\citep{Skill75a}, (note that here $k=k_{\parallel})$. We have included
only the linear wave damping $\Gamma$ and we will return to the possible
role of nonlinear saturation effects later. We assume that $\partial P_{{\rm CR}}/\partial z\le0$
at all times, so that only the forward propagating waves are unstable.
The latter inequality is ensured by the formulation of initial value
problem in each of the following two settings mentioned earlier in
this section. In the first setting, the initial distribution of the
CRC is symmetric with respect to $z=0$, so we can consider their
escape into the half-space $z>0$. The second setting is when the
cloud is limited from the left by a reflecting wall (CD). The appropriate
boundary condition is $\partial P_{{\rm CR}}/\partial z=0$ at $z=0$
in both cases.

Restricting our consideration to the case of coordinate-independent
damping rate $\Gamma$, we obtain the following ('quasilinear') integral
of the system of Equations (\ref{eq:dPdt}) and (\ref{eq:dIdt}):

\begin{equation}
P_{{\rm CR}}\left(z,t\right)=P_{{\rm CR}0}\left(z^{\prime}\right)-\frac{\kappa_{{\rm B}}}{C_{{\rm A}}}\frac{\partial}{\partial z}\ln\frac{I\left(z,t\right)}{I_{0}\left(z^{\prime}\right)}\label{eq:QLint}
\end{equation}
Here $P_{{\rm CR}0}\left(z\right)$ and $I_{0}\left(z\right)$ are
the initial distributions of the CR partial pressure and the wave
energy density, respectively, and $z^{\prime}=z-\left(U+C_{{\rm A}}\right)t$. 

Using the quasi-linear integral, the system of Equations (\ref{eq:dPdt})
and (\ref{eq:dIdt}) can be reduced to one nonlinear convection-diffusion
equation for \eg wave intensity $I\left(z,t\right)$. We will use
dimensionless variables measuring the distance $z$ in units of $a_{\parallel}$,
which is the initial size of the CRC along the field line. The time
unit is then $a/C_{{\rm A}}$. Note, however, that, as the acceleration
is assumed to be inactive, the particle momentum $p$ enters the problem
only as a parameter but, the initial scale-height of the CRC $a$
generally depends on $p$. It is also convenient to introduce a new
variable for the wave energy density

\begin{equation}
W=\frac{C_{{\rm A}}a\left(p\right)}{\kappa_{{\rm B}}\left(p\right)}I\label{eq:Wnormaliz}
\end{equation}
and similar relations for $I_{0}$ and $W_{0}$. The equation for
$W$ then takes the following form

\begin{equation}
\frac{\partial W}{\partial t}-\frac{\partial}{\partial z}\frac{1}{W}\frac{\partial W}{\partial z}=-\frac{\partial}{\partial z}\mathcal{P}_{0}\left(z\right)\label{eq:dwdt}
\end{equation}
We have neglected the wave propagation along the characteristics given
by eq.(\ref{eq:charct}) and the wave damping, assuming that these
processes are slower than the CRC diffusion: $U,C_{{\rm A}},a\Gamma\ll\kappa_{{\rm B}}/aI\sim cr_{{\rm g}}/aI$.
When the linear damping is important (e.g. in the case of Goldreich-Shridhar
cascade, \citealp{FarmerGoldr04}), it can be easily incorporated
into the current treatment by a simple change of variables indicated
in Sec.\ref{sec:Comparison-with-other}. The function $\mathcal{P}$
is defined similarly to $W$ above

\begin{equation}
\mathcal{P}=\frac{C_{{\rm A}}a}{\kappa_{{\rm B}}\left(p\right)}P_{{\rm CR}}\label{eq:Pnormalization}
\end{equation}
and, again, the index $0$ refers in Equation (\ref{eq:dwdt}) to
the initial CR distribution $P_{{\rm CR}0}\left(z\right)$. Thus,
according to eq.(\ref{eq:QLint}), for the dimensionless CR partial
pressure we have

\begin{equation}
\mathcal{P}=\mathcal{P}_{0}-\frac{\partial}{\partial z}\ln\frac{W}{W_{0}}\label{eq:PofW}
\end{equation}
This quantity is governed by the equation

\begin{equation}
\frac{\partial\mathcal{P}}{\partial t}=\frac{\partial}{\partial z}\frac{1}{W}\frac{\partial\mathcal{P}}{\partial z},\label{eq:PDifEq}
\end{equation}
but its solution can be written down using the integral given by equation
(\ref{eq:PofW}), after the solution to Equation (\ref{eq:dwdt})
is obtained.

While letting $C_{{\rm A}}\to0$ in the wave-particle collective propagation,
we utilize the finiteness of $C_{{\rm A}}$ in determining the boundary
condition for eq.(\ref{eq:dwdt}) at $z=0$ as follows. Returning
from the nonlinear diffusion equation eq.(\ref{eq:dwdt}) to eq.(\ref{eq:dIdt}),
one may see that the wave energy does not actually ``diffuse'' but
it is generated locally by particles that diffuse. Had the neglected
wave diffusionspread the waves to the point of their stability viz.
$z=0$, they would be convected away from it by virtue of $C_{{\rm A}}>0$.
Recalling the symmetry (reflection) boundary condition $\partial P_{{\rm CR}}/\partial z=0$
at $z=0$ (no wave generation), we thus set a fixed boundary value
$W\left(0,t\right)=W_{0}$, where $W_{0}$ is an initial (small) wave
noise. It is instructive to investigate the case in which the initial
noise is the same throughout the entire half-space $z>0$ (this limitation
can be straightforwardly relaxed), so that the waves will be generated
entirely by escaping particles, thus emphasizing the self-confinement.
The second boundary condition is set by $W\to W_{0}$ for $z\to\infty$,
and all $t<\infty$. Note that in general, $W_{0}=W_{0}\left(p\right)$.
These conditions determine the boundary value problem given by eq.(\ref{eq:dwdt})
completely. However, we are interested primarily in diffusion of CRs
outside the region of their initial localization, viz. at $z>1$,
where the source term in this equation vanishes. Therefore, the problem
given by eq.(\ref{eq:dwdt}) can be split into two separate problems,
one in $0\le z\le1$ and another in $1<z<\infty$ domain with the
following junction conditions

\begin{equation}
\left.\left.\frac{\partial}{\partial z}\ln W\right|_{1-}^{1+}=W\right|_{1-}^{1+}=0.\label{eq:junction}
\end{equation}
These are the continuity conditions for both the wave energy density
and pressure across the edge of the initial CR localization.

\subsection{Self-similar solution outside the region of initial CR localization\label{sub:Self-similar-solution-outside}}

The nonlinear boundary value problem in the region $z>1$ given by
Equations (\ref{eq:dwdt}) and (\ref{eq:junction}) describes the
CR propagation outside the region of their initial localization. This
problem can be solved exactly using the following self-similar substitution:

\begin{equation}
W=\frac{1}{t^{\alpha}}w\left(\zeta\right),\;\;\;{\rm where}\;\zeta=\frac{z}{t^{\beta}}\label{eq:self-s-subst}
\end{equation}
Submitting this to eq.(\ref{eq:dwdt}) with $\mathcal{P}_{0}=0$ yields
$\alpha=2\beta-1$. The boundary condition at infinity ($W\to W_{0}\neq0,\; z\to\infty$),
on the other hand, requires $\alpha=0$, so that the equation for
$w$ reads

\begin{equation}
\frac{d}{d\zeta}\frac{1}{w}\frac{dw}{d\zeta}+\frac{\zeta}{2}\frac{dw}{d\zeta}=0\label{eq:w}
\end{equation}
with $\zeta=z/\sqrt{t}$. It is interesting to observe that this equation
has a simple special solution $w=2/\zeta^{2}=2t/z^{2}$. This solution
describes a stationary particle distribution $\mathcal{P}\propto1/z$
that is obviously related to the \citet{Bell78} asymptotic particle
distribution in self-excited waves far away from a shock. In both
cases, this is a singular limit of the problem as it implies a zero
background turbulence level, $W_{0}=0$ and thus the number of particles
in $z>0$ half-space being infinite \citet{LagCes83a,Drury83}. Physically,
this solution is different from what we will find below in that it
requires a permanent source of CR at the origin so that their flux
steadily drives waves that linearly grow in time.%
\footnote{There is no convection with the flow towards the origin ($U=0$) in
our case, which makes the solution unsteady, in contrast to the corresponding
DSA problem \citep{Bell78} where the solution is steady because of
the convection, but the number of particles upstream is still infinite
in both cases.%
} As we shall see, this simple special solution is an important singular
limit of a more general and completely regular solution.

\begin{figure}
\includegraphics[bb=0bp 0bp 612bp 792bp,scale=0.7, angle=270]{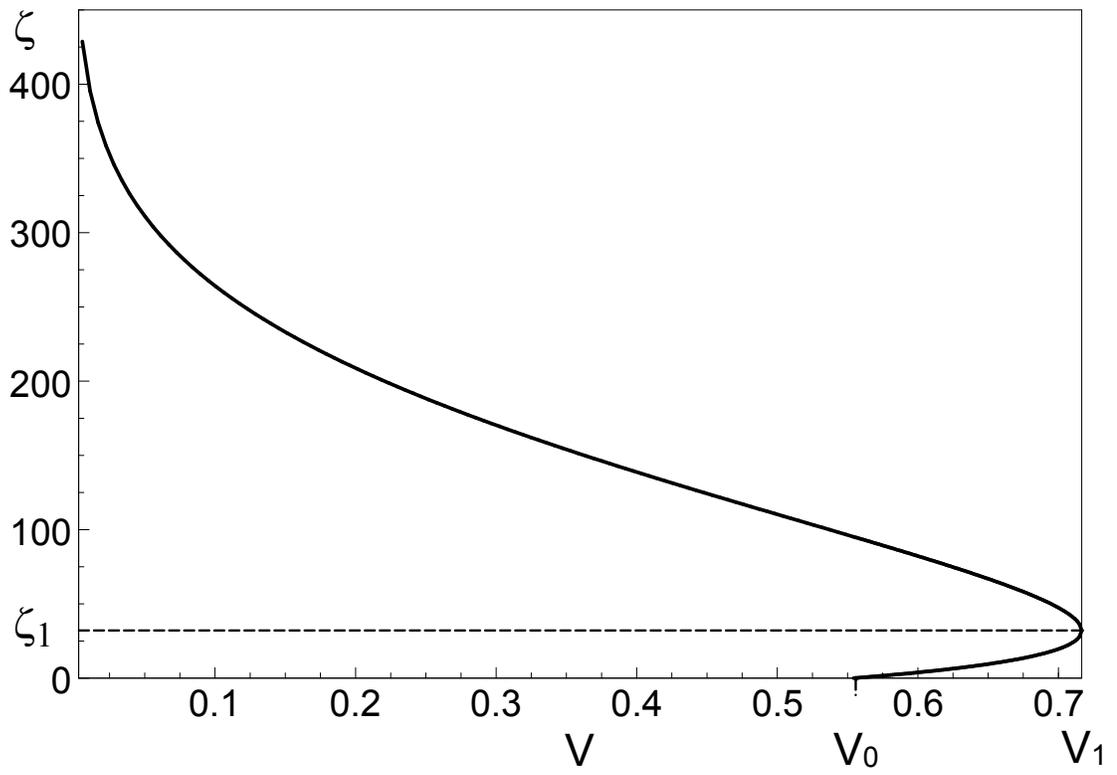}

\caption{Two branches of function $\zeta\left(V\right)$, eq.(\ref{eq:zetaAp}),
depicted for $\varepsilon=0.3$. Note that $R\left(V=0\right)=\infty$,
$R\left(V_{0}\right)=V_{0}/4,$ $R\left(V_{1}\right)=0$ and $\zeta\left(V_{0}\right)=0$
on the lower branch of $\zeta\left(V\right)$.\label{fig:FunctionRofV}}
\end{figure}

\begin{figure}
\includegraphics[clip,scale=0.7, angle=270]{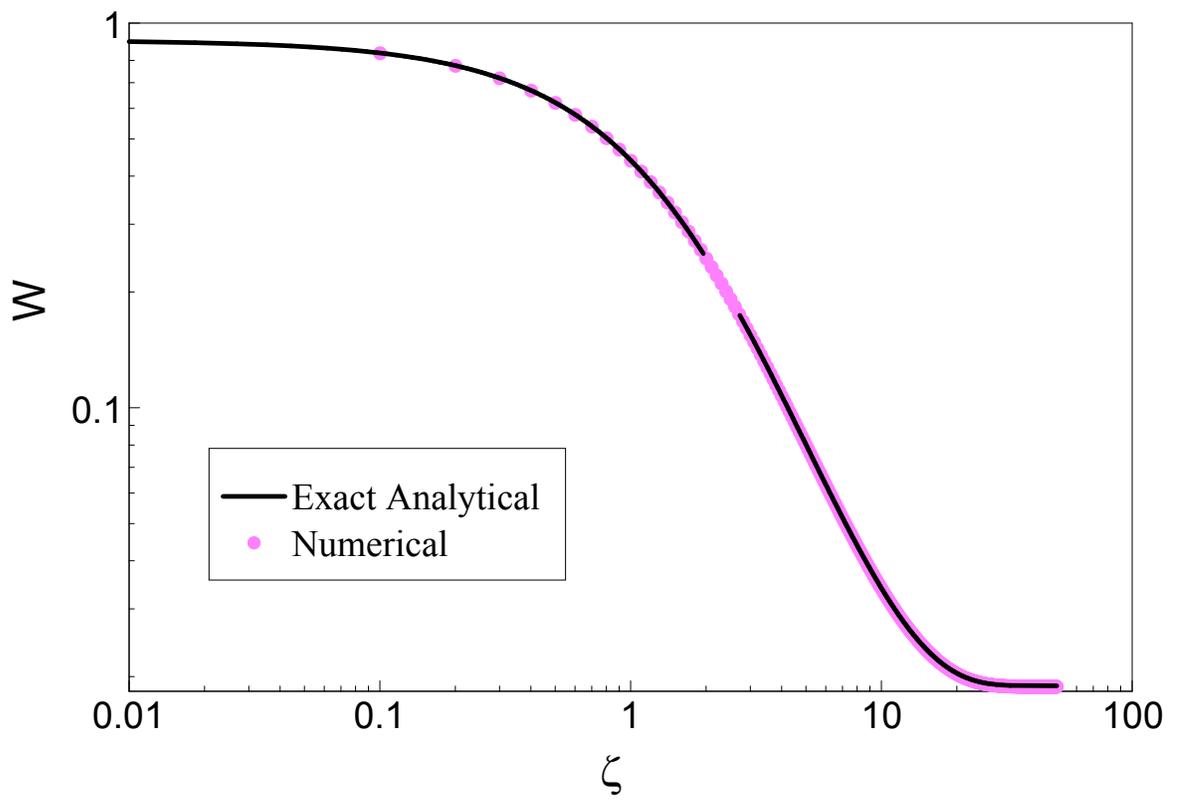}

\caption{Analytic vs numerical solution of eq.(\ref{eq:dwdzeta}). The gap
in the analytical curve encloses the branching point of the solution
at $V=V_{1}$, eq.(\ref{eq:wofVfull}). $w_{0}=0.19$, $w_{m}=0.9$\label{fig:Analytic-vs-numerical} }
\end{figure}

The general solution to eq.(\ref{eq:w}) can be found in quadratures
by swapping $\zeta$ and $w$ as dependent and independent variables
and introducing an auxiliary function $V\left(w\right)$ by the following
substitution

\begin{equation}
\frac{dw}{d\zeta}=-\sqrt{2}w^{3/2}V\left(w\right).\label{eq:dwdzeta}
\end{equation}
The equation for $V\left(w\right)$ takes the following form

\begin{equation}
w\frac{d}{dw}w\frac{dV}{dw}+\frac{1}{4}\left(\frac{1}{V}-V\right)=0\label{eq:dVdw}
\end{equation}
This equation can be easily integrated, so that the function $w\left(\zeta\right)$
can be subsequently found from eq.(\ref{eq:dwdzeta}). The first integral
of eq.(\ref{eq:dVdw}) is as follows

\begin{equation}
\left(\frac{\partial V}{\partial\ln w}\right)^{2}=R\left(V\right)\equiv\frac{1}{4}\left(V^{2}-2\ln V-q\right)\label{eq:FirstInt}
\end{equation}
where $q$ is an integration constant. From eq.(\ref{eq:dwdzeta})
and from the boundary condition $w\to W_{0}$ at $\zeta\to\infty$
we infer that $V\left(W_{0}\right)=0$, so that from eq.(\ref{eq:FirstInt})
we find

\begin{equation}
w=W_{0}e^{\int_{0}^{V}\frac{dV^{\prime}}{\sqrt{R\left(V^{\prime}\right)}}}\label{eq:wOfV}
\end{equation}
The function $w\left(\zeta\right)$ is then determined by the following
relation for $\zeta\left(V\right)$, eq.(\ref{eq:dwdzeta}):

\begin{equation}
\zeta=-\frac{1}{\sqrt{2W_{0}}}\intop\frac{dV}{V\sqrt{R\left(V\right)}}e^{-\frac{1}{2}\int_{0}^{V}\frac{dV^{\prime}}{\sqrt{R\left(V^{\prime}\right)}}}\label{eq:zetaOfV}
\end{equation}
Using the identity $1/V=V-2dR/dV$ and integrating eq.(\ref{eq:zetaOfV})
by parts, after some manipulations the last equation simplifies considerably:

\begin{equation}
\zeta=\sqrt{\frac{2}{w}}\left(2\sqrt{R}+V\right)\label{eq:zetaofV1}
\end{equation}
It is useful to introduce an auxiliary constant $V_{1}$ related to
the constant $q$ and being the smaller of the two roots of $R\left(V\right)$,
that is $R\left(V_{1}\right)=0$. The requirement of a real zero of
the function $R\left(V\right)$ ensures mapping two 'copies' of the
domain of $V$ onto the full range $0<\zeta<\infty$, whereby $\zeta\to\infty$
for $V\to0$. Indeed, $\zeta\left(V\right)$ in eq.(\ref{eq:zetaofV1})
diverges at $V=0$ as $\sim\sqrt{-\ln V}$. However, increasing $V$
from $V=0$ to $V=V_{1}$ does not cover the full range $0<\zeta<\infty$
yet, as may be seen from eq.(\ref{eq:zetaofV1}); $\zeta$ decreases
from $\infty$ only to $\zeta_{1}\equiv V_{1}\sqrt{2/w\left(V_{1}\right)}>0$.
To continue the integral curve to $\zeta=0$, it is necessary to switch
branches of $\sqrt{R}$ at $V=V_{1}$ in eqs.(\ref{eq:wOfV}) and
(\ref{eq:zetaofV1}) and so continue the integral in eq.(\ref{eq:wOfV})
back to $V<V_{1}$ (along the second copy of the $V$-domain). Decreasing
then $V$ from $V=V_{1}$ to $V=V_{0}\equiv\exp\left(-q/2\right)$
brings the integral curve to the point $\zeta=0$, $w=w_{{\rm max}}=w\left(\zeta=0\right)$,
Fig.\ref{fig:FunctionRofV}. The explicit formal representation of
the general solution of eq.(\ref{eq:w}) is given in Appendix \ref{sec:Details-of-self-similar}
along with a numerical example, Fig.\ref{fig:Analytic-vs-numerical}.
Equations (\ref{eq:wOfV}) and (\ref{eq:zetaofV1}) determine the
solution $w\left(\zeta\right)$, where the integration constant $V_{0}$
(or $q)$ should be obtained from the matching condition, eq.(\ref{eq:junction}).
This can be done by considering eq.(\ref{eq:dwdt}) inside the region
of initial CR localization, i.e. $0<z<1$, where its r.h.s is nonzero.

\subsection{Solution inside the initial CR cloud\label{sub:Solution-inside-of}}

Given an initial distribution $\mathcal{P}_{0}\left(z\right)$, equation
(\ref{eq:dwdt}) can always be integrated numerically in the finite
domain $0<z<1$, so that the full solution will be obtained from the
boundary conditions and from the results of the previous section.
However, according to Equation (\ref{eq:PDifEq}), already for $t\gtrsim W_{0}$,
where $W_{0}\ll1$, the initial profile $\mathcal{P}_{0}\left(z\right)$
will be redistributed over the unity interval in such a way as to
approach a quasi-steady state in which the flux $W_{0}^{-1}\partial W/\partial z$
in Equation (\ref{eq:dwdt}) through $z=0$ will be balanced by the
source integral, $\mathcal{P}_{0}\left(0\right)$ (recall that $\mathcal{P}_{0}\left(1\right)=0$).
The particle flux through the $z=1$ boundary decays as $t^{-1/2}$
with time (since $\left|\partial w/\partial\zeta\right|<\infty$ at
$\zeta\to0+$, see the preceding section or Equation (\ref{eq:Boft})
below). Therefore, for the self-similar stage of the cloud relaxation
we can write

\begin{equation}
\ln\frac{W}{W_{0}}=\intop_{0}^{z}\mathcal{P}_{0}dz-B\left(t\right)z\label{eq:lnW}
\end{equation}
where the ``integration constant'' $B$ depends slowly (in the above
sense) on time. Using the first matching condition in Equation (\ref{eq:junction}),
\ie the CR pressure continuity, we can specify $B\left(t\right)$
as follows: $B\propto t^{-1/2}\to0$ as $t\to\infty$ (see Equation
{[}\ref{eq:Boft}{]} below). This determines the self-similar (outer,
$z>1$) solution by the second matching condition in Equation (\ref{eq:junction}):

\begin{equation}
w\left(0\right)\equiv\lim_{t\to\infty}w\left(\frac{1}{\sqrt{t}}\right)=w_{{\rm max}}=W_{0}e^{\Pi}\label{eq:wmaxdef}
\end{equation}
Here we have introduced the integrated partial pressure as follows

\begin{equation}
\Pi=\int_{0}^{1}\mathcal{P}_{0}dz.\label{eq:Pidef1}
\end{equation}
The function $B\left(t\right)$ and thus the internal solution $W\left(z,t\right)$
in Equation (\ref{eq:lnW}) may be obtained using this equation, the
first matching condition in Equation (\ref{eq:junction}), and Equations
(\ref{eq:dwdzeta}) 

\begin{equation}
B\left(t\right)=-\left.\frac{1}{\sqrt{t}}\frac{\partial}{\partial\zeta}\ln w\right|_{\zeta=1/\sqrt{t}}=\left.\sqrt{2w/t}V\right|_{\zeta=1/\sqrt{t}}\label{eq:Boft}
\end{equation}
Note that the particle pressure at $z>1$ is completely determined
by the turbulence level $w$, eq.(\ref{eq:PofW}). 

Now we can determine the integration constant $q$ introduced in the
preceding section. From Equations (\ref{eq:wmaxdef}) and (\ref{eq:wmaxAp})
we obtain the following equation for $q$

\begin{equation}
\int_{0}^{V_{1}}dV/\sqrt{R\left(V\right)}+\int_{V_{0}}^{V_{1}}dV/\sqrt{R\left(V\right)}=\Pi,\label{eq:Pidef2}
\end{equation}
where $R\left(V_{1}\right)=0$, $R\left(V_{0}\right)=-V_{0}/2$, while
$R\left(V\right)$ is given by Equation (\ref{eq:FirstInt}). In the
most interesting case $q\approx1$, it is convenient to use the constant
$\varepsilon^{2}=\left(q-1\right)/2$ in place of $q$. In this case
$\Pi\gg1$, $\varepsilon\ll1$, $V_{1}\approx1-\varepsilon$ and

\[
\Pi=2\intop_{0}^{V_{1}}\frac{dV}{\sqrt{R}}\approx2^{3/2}\intop_{0}^{1-\varepsilon}\frac{dV}{\sqrt{\left(V-1\right)^{2}-\varepsilon^{2}}}\approx2^{3/2}\ln\frac{2}{\varepsilon}+\mathcal{O}\left(1\right),
\]
so that the turning point of the solution at $V=V_{1}$ approaches
the critical point $V=1$, where $R$ has a minimum (Fig.\ref{fig:FunctionRofV}):

\[
V_{1}=1-2e^{-\Pi/2\sqrt{2}}.
\]
For $\Pi\ll1$, we can write instead, $V_{1}\approx V_{0}\left(1+V_{0}^{2}/2\right)$
where $V_{0}=\exp\left(-q/2\right)$. 

To conclude this section, the self-similar expansion of the CRC described
by Equations (\ref{eq:dwdt}-\ref{eq:PDifEq}) is controlled by two
parameters. One parameter is the background turbulence level $W_{0}\left(p\right)$
in the media into which the cloud expands. The second parameter is
the integrated pressure of the cloud, $\Pi$. Although the initial
wave energy density inside the cloud ($z<1$) is likely to be higher
than $W_{0}$, we have adopted the background value $W_{0}$ also
inside the cloud for simplicity, as this should not influence the
self-similar CR propagation outside the cloud. In addition, efficient
wave generation by a dense expanding CRC, renders the initial value
for the wave energy density inside the CRC unimportant.

\section{Analysis of the solution\label{sec:Analysis-of-the}}

Once we have the solutions outside and inside the region of initial
CR localization, we may precisely calculate how fast particles escape
from this region. Two convenient characteristics of the escape process
are the half-life time and the width of the CR distribution. We begin
our analysis of the solution from computing these simple quantities
and turn to the details of the CR escape afterward.

\subsection{Half-life time and the width of the CR cloud\label{sub:Half-life-and-the}}

The most concise characterization of the escape may be given by looking
at the following two parts of the (conserved) integral particle pressure:

\[
\Pi=\Pi_{0}+\Pi_{1}={\rm const}
\]
where

\[
\Pi_{0}=\intop_{0}^{1}\mathcal{P}dz\;\;\;\;{\rm }{\rm and}\;\;\;\;\Pi_{1}=\intop_{1}^{\infty}\mathcal{P}dz
\]
refer to the regions inside and outside of the initial CRC, respectively.
Recall that at $t=0$, $\Pi_{0}=\Pi$ and $\Pi_{1}=0$, while at $t=\infty$,
an opposite CR distribution is reached: $\Pi_{0}=0$ and $\Pi_{1}=\Pi$.
Note that Equation (\ref{eq:wmaxdef}) specifies the total work ultimately
done by particles on waves, while they diffuse from $0<z<1$ to $1\le z<\infty$.
As $\Pi$ is conserved, it is natural to define the CR confinement
time as the time at which $\Pi_{0}$ ($\Pi_{1}$) drops (raises) to
a half of $\Pi$. Substituting $W$ from Equation (\ref{eq:self-s-subst})
into Equation (\ref{eq:PofW}), we obtain

\[
\Pi_{1}\left(t\right)=\left.\ln\frac{w}{W_{0}}\right|_{\zeta=1/\sqrt{t}}
\]
For the half-life time $t=t_{{\rm 1/2}}$ we may use the following
equations: 

\begin{equation}
\Pi_{1}\left(t_{1/2}\right)=\Pi_{0}\left(t_{1/2}\right)=\frac{1}{2}\Pi.\label{eq:HalfLifetDef}
\end{equation}
In the simple test particle case $(\Pi\ll1$), the half-life time
amounts to (see Appendix \ref{sec:Half-life-of-CR}) 

\begin{equation}
t_{1/2}\approx\frac{1}{4\sigma}W_{0}\label{eq:thalflin}
\end{equation}
 with $\sigma\approx0.23$. In the opposite case $\Pi\gg1$, for $t_{1/2}$
we obtain

\begin{equation}
t_{1/2}\approx t_{1}=1/\zeta_{1}^{2}\equiv\frac{w_{1}}{2V_{1}^{2}}\approx\frac{w_{1}}{2}\label{eq:thalfNL}
\end{equation}
(see Equation {[}\ref{eq:zetaofV1}{]}). Note that for $\Pi\gg1$,
$V_{1}\approx V_{1/2}$, (see Appendix \ref{sec:Half-life-of-CR})
and for the half-life time we obtain

\begin{equation}
t_{1/2}=\frac{1}{2}W_{0}e^{\Pi/2}\label{eq:thalfLINNL}
\end{equation}
It is clear that the nonlinear delay factor $\exp\left(\Pi/2\right)$
slows down the escape considerably, compared to the test particle
solution.

Another important characteristics of the particle escape is the spatial
width of their distribution. Formally, the self-similar solution is
scale-invariant $\sim2/\zeta$, for $\Pi\gg1$ over the most part
of the spatial distribution of the partial pressure (see below). Therefore,
to characterize the width we use the point $\zeta_{1/2}$, related
to the half time $t_{1/2}$ as follows:

\[
\zeta_{1/2}=\frac{1}{\sqrt{t_{1/2}}}=\sqrt{\frac{2}{W_{0}}}e^{-\Pi/4}
\]
In physical coordinate $z$, this point moves outwards as $z_{1/2}=\zeta_{1/2}\sqrt{t}$
starting from $z=1$ at $t=t_{1/2}$. The expansion rate of a dense
CRC ($\Pi\gg1$) is thus exponentially low.

\subsection{Spatial distribution of the CR cloud\label{sub:SpatialProfileOfP}}

\begin{figure}
\includegraphics[scale=0.7, angle=270]{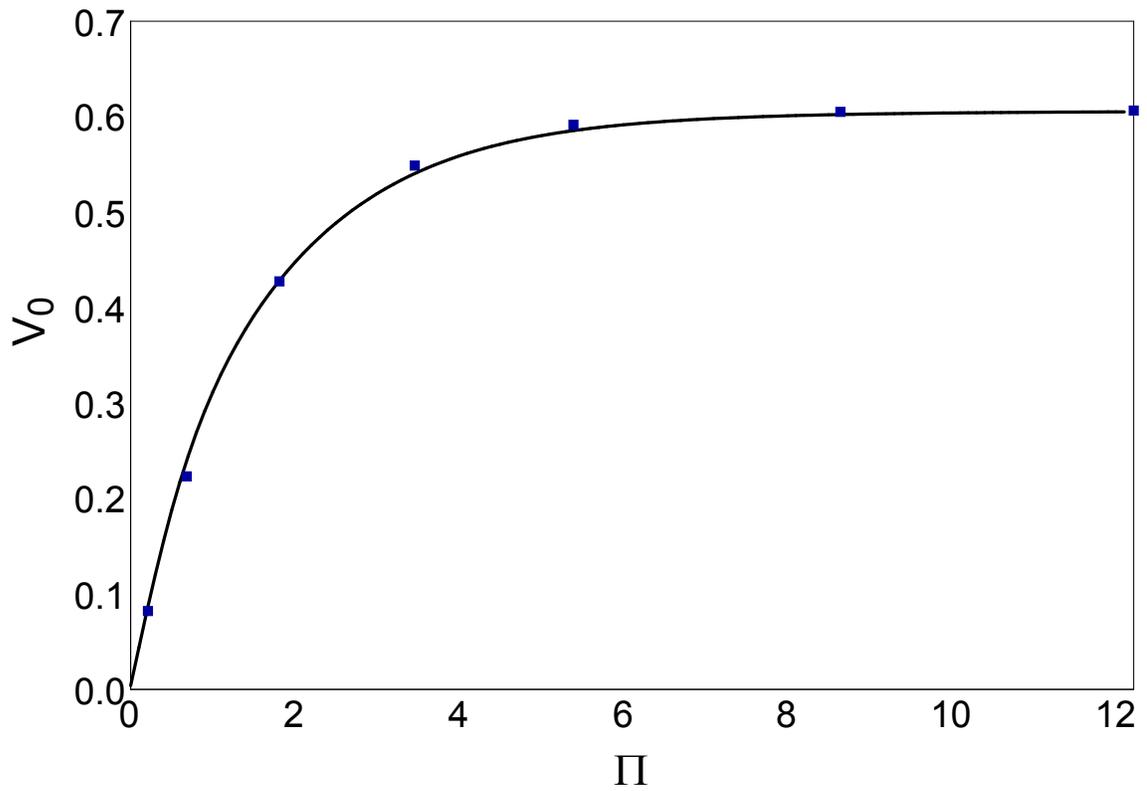}

\caption{Squares: Function $V_{0}\left(\Pi\right)$, obtained from Equation
(\ref{eq:Pidef2}). Line: interpolation given by Equation (\ref{eq:V0Interpol}).\label{fig:V0Interpol}}
\end{figure}

Considering the spatial distribution of the spreading CRC in detail,
it is convenient to start with the region far away from the source,
where $W\to W_{0}$. This asymptotic behavior corresponds to the case
$V\ll1$. Introducing an auxiliary function $U\left(\zeta\right)$

\begin{equation}
U=-\frac{\partial}{\partial\zeta}\ln w=\sqrt{2w}V\label{eq:UofwandV-1}
\end{equation}
that is related to the particle pressure through $\mathcal{P}=U\left(\zeta\right)/\sqrt{t}$,
and using Equations (\ref{eq:FirstInt}) and (\ref{eq:zetaofV1}),
we obtain for $U$ the following expression

\begin{equation}
U\left(\zeta\right)=\sqrt{2W_{0}}V_{0}e^{-W_{0}\zeta^{2}/4}\label{eq:ULargeZeta}
\end{equation}
Therefore, the asymptotic CR pressure depends on the following two
parameters: the ISM background diffusivity $W_{0}^{-1}\left(p\right)\equiv W^{-1}\left(p,z=\infty\right)$
(see Equation {[}\ref{eq:PDifEq}{]}) and on the CR source pressure
$\Pi\left(p\right)$, but only through the parameter $V_{0}=\exp\left(-q/2\right)$.
The latter grows linearly with $\Pi\ll1$ but it saturates when $\Pi\gg1$:

\begin{equation}
V_{0}\approx\left\{ \begin{array}{cc}
\frac{1}{\sqrt{2\pi}}\Pi, & \Pi\ll1\\
\exp\left(-\frac{1}{2}-4e^{-\Pi/\sqrt{2}}\right), & \Pi\gg1
\end{array}\right.\label{eq:V0cases}
\end{equation}
For practical calculations, it is more convenient to combine both
asymptotic regimes into the following simple interpolation formula:

\begin{equation}
V_{0}=\left[\left(\sqrt{2\pi}/\Pi\right)^{5/2}+\left(\sqrt{e}+\frac{3}{2}e^{-\Pi/\sqrt{2}}\right)^{5/2}\right]^{-2/5}\label{eq:V0Interpol}
\end{equation}
which not only recovers both $\Pi\ll1$ and $\Pi\gg1$ regimes but
also reproduces the transition zone accurately. The latter property
is achieved by the choice of numerical parameters of the interpolation,
i.e., the powers $5/2$, 3/2 and $-2/5$. The quality of the interpolation
may be seen from Fig.\ref{fig:V0Interpol}, where the formula in Equation
(\ref{eq:V0Interpol}) is shown against exact numerical points calculated
from Equation (\ref{eq:Pidef2}). The partial pressure $\mathcal{P}\left(z,t\right)$
is then given by 

\begin{equation}
\mathcal{P}=U/\sqrt{t}=\sqrt{2W_{0}/t}V_{0}\left(\Pi\right)e^{-W_{0}z^{2}/4t}\label{eq:Pasympt}
\end{equation}
Note, that for strong sources ($\Pi\gg1$) the CR density at large
distances becomes independent of the source strength. This is the
regime of an efficient ablative escape suppression in which a small
(and $\Pi$-independent!) number of leaking particles leave behind
enough Alfv\'en fluctuations to limit the leakage of particles remaining
in the source to exactly the rate given by Equation (\ref{eq:Pasympt}).

The spatial distribution of the CRs becomes even more universal closer
to the origin where it falls off as $\mathcal{P}\approx2/z$ before
it turns into an innermost flat-top part of the entire distribution
for $\zeta^{2}<D_{{\rm NL}}\left(p\right)$. We have introduced the
``self-confinement'' diffusion coefficient $D_{{\rm NL}}$ as

\begin{equation}
D_{{\rm NL}}=\frac{2}{V_{0}^{2}}D_{{\rm ISM}}e^{-\Pi}=\frac{2}{V_{0}^{2}w_{{\rm max}}}\label{eq:DNLdef}
\end{equation}
with $D_{{\rm ISM}}=W_{0}^{-1}$. To obtain this intermediate ($D_{{\rm NL}}<\zeta^{2}<D_{{\rm ISM}}$)
part of the CR spatial distribution we expand the analytic solution
given by Equations (\ref{eq:wOfV}) and (\ref{eq:zetaofV1}) in small
$V_{1}-V\ll1$. Using the expression for the self-similar CR pressure
$U$ from Equation (\ref{eq:UofwandV-1}) we obtain the following
expansion near $\zeta=\zeta_{1}\equiv\zeta\left(V_{1}\right)$:

\begin{equation}
U=\frac{2}{\zeta}\left[1-\frac{3\epsilon}{2}\left(\frac{2}{w_{1}}\right)^{2^{-3/2}}\zeta^{-\sqrt{2}}-\frac{\epsilon}{2}\left(\frac{w_{1}}{2}\right)^{2^{-3/2}}\zeta^{\sqrt{2}}\dots\right],\label{eq:UintermExp}
\end{equation}
where $w_{1}=2V_{1}^{2}/\zeta_{1}^{2}$, and $\epsilon\equiv\left(q-1\right)/2$.
Note that while the solution has a branching point in auxiliary variable
$V$ at $V=V_{1}$, it is regular and single-valued function of the
physical variable $\zeta$, throughout the entire half space $\zeta>0$,
including the branching point of $\zeta$$\left(V\right)$ at $V=V_{1}$,
as it should be. We also notice that while $\zeta$ decreases starting
from $\zeta_{1}$, the solution grows slower than $1/\zeta$, leveling
off towards the origin. The above expansion, however, becomes inaccurate
for smaller $\zeta$ and we alter it below.

\begin{figure}
\includegraphics[scale=0.7, angle=270]{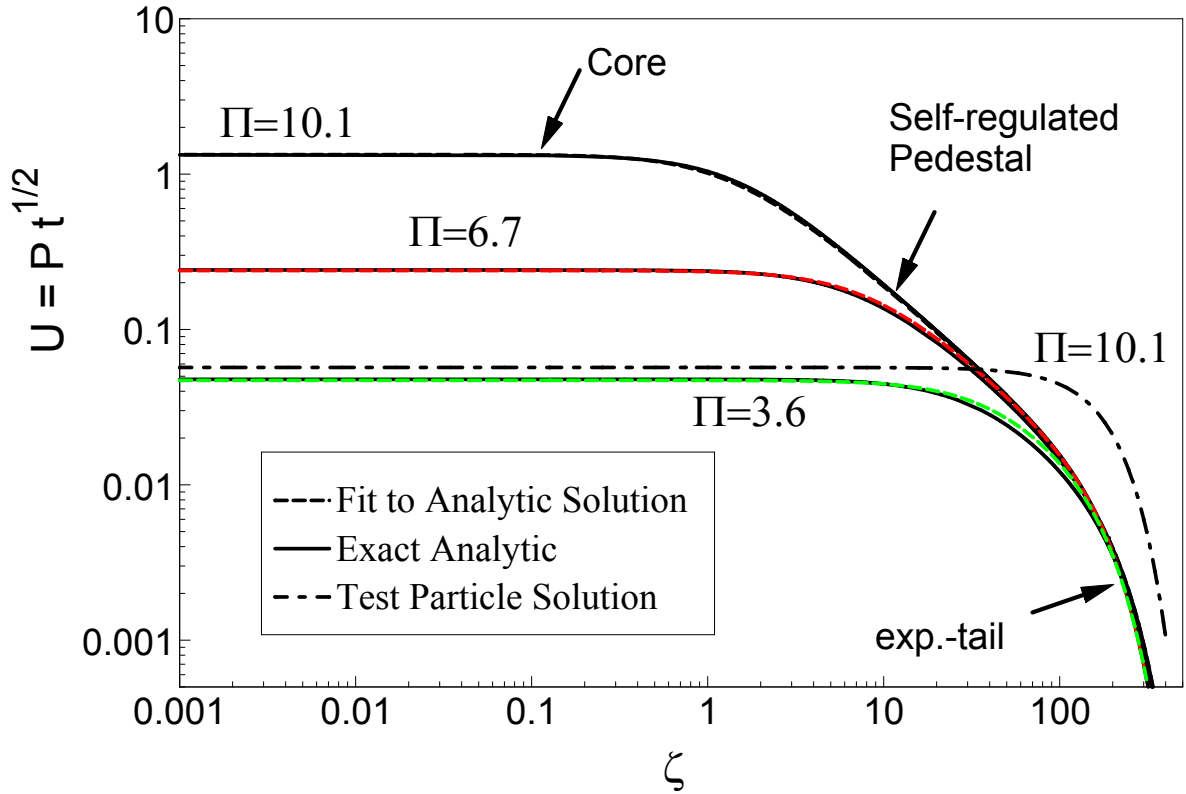}

\caption{Spatial distribution of CR partial pressure (as a function of $\zeta=z/\sqrt{t}$,
multiplied by $\sqrt{t}$) shown for integrated values of this quantity
$\Pi=3.6;$ 6.7; 10.1 and for for the background wave amplitude $W_{0}=10^{-4}.$
Exact analytic solutions are shown with the solid lines while the
interpolations given by Equation (\ref{eq:Ufit}) are shown with the
dashed lines. For comparison, a formal linear solution for $\Pi=10.1$
is also shown with the dot-dashed line. Note the three characteristic
zones of the CR confinement: the innermost flat top core, the scale
invariant ($1/\zeta$) pedestal, and the exponential decay zone. \label{fig:SpatialProfileFits}}
\end{figure}

Now we turn to this innermost part of the distribution where the particle
diffusion coefficient is most strongly decreased. It is convenient
to expand the solution in a series in $\xi=V/V_{0}-1\ll1$. Using
the representation of $w$ given by Equation (\ref{eq:Apwmax}), we
obtain

\[
w=w_{{\rm max}}\left[\frac{\xi-a+\sqrt{\xi^{2}-2a\xi+b^{2}}}{b-a}\right]^{-2b}
\]
where $a=\left(1-V_{0}^{2}\right)/\left(1+V_{0}^{2}\right)$ and $b=V_{0}/\sqrt{1+V_{0}^{2}}$.
Then, using the expression (\ref{eq:zetaAp}) for small $\zeta$ and
Equation (\ref{eq:UofwandV-1}), we obtain for $U\left(\zeta\right)=\mathcal{P}\sqrt{t}$
the following simple result

\begin{equation}
U\approx\frac{\sqrt{2w_{{\rm max}}}V_{0}}{1+w_{{\rm max}}\zeta^{2}/4}\approx\sqrt{2w_{{\rm max}}}V_{0}e^{-w_{{\rm max}}\zeta^{2}/4},\label{eq:UsmallZeta}
\end{equation}
valid for $w_{{\rm max}}\zeta^{2}\lesssim1$. Note that we have rewritten,
with the same accuracy, the denominator in the standard ``diffusion''
form, $\exp\left(-w_{{\rm max}}\zeta^{2}/4\right)$, which shows that
the CR diffusivity is diminished by a factor $W_{0}/w_{{\rm max}}=\exp\left(-\Pi\right)\ll1$
compared to its background level $W_{0}^{-1}$. If, however, $w_{{\rm max}}\gtrsim aC_{{\rm A}}/\kappa_{{\rm B}}$
(see Equation {[}\ref{eq:Wnormaliz}{]}), a better approximation would
be to simply replace the diffusion coefficient by its Bohm value,
as the neglected nonlinear wave interactions (see Sec.\ref{sec:Comparison-with-other})
should render $\delta B\sim B_{0}$. The result in Equation (\ref{eq:UsmallZeta})
should then be replaced by 
\begin{equation}
U=\sqrt{2aC_{{\rm A}}/\kappa_{{\rm B}}}V_{0}\left(\Pi\right)e^{-aC_{{\rm A}}z^{2}/4\kappa_{{\rm B}}t}\label{eq:UBohm}
\end{equation}
Eqs.(\ref{eq:ULargeZeta}), (\ref{eq:UintermExp}) and (\ref{eq:UsmallZeta})
provide the \emph{explicit }asymptotic representation of the exact
\emph{implicit }solution given by eqs.(\ref{eq:wOfV}) and (\ref{eq:zetaofV1}).
This representation is rigorous but somewhat impractical, so we provide,
again, an interpolation formula that accurately describes the solution
$U\left(\zeta\right)=\mathcal{P}\sqrt{t}$ in the entire range of
$0<\zeta<\infty$, Fig.\ref{fig:SpatialProfileFits}

\begin{equation}
U=2\left[\zeta^{5/3}+\left(D_{{\rm NL}}\right)^{5/6}\right]^{-3/5}e^{-W_{0}\zeta^{2}/4}\label{eq:Ufit}
\end{equation}
where $D_{{\rm NL}}\left(\Pi\right)$ is defined in eq.(\ref{eq:DNLdef})
with $V_{0}$ given by the interpolation in Equation (\ref{eq:V0Interpol}).
The last formula recovers the limiting cases given by Equations (\ref{eq:ULargeZeta}),
(\ref{eq:UintermExp}) and (\ref{eq:UsmallZeta}) except a slight
modification of Equation (\ref{eq:ULargeZeta}) at large $\zeta$,
so that an extra $1/\zeta$-factor is accepted in the interests of
parametrization simplicity. According to Fig.\ref{fig:SpatialProfileFits},
however, this deviation from the correct asymptotic expansion of the
exponential tail of the distribution is not really noticeable. 

The overall profile of the partial pressure presented in log-log format
(Fig.\ref{fig:SpatialProfileFits}), unveils the CR escape as a highly
structured process. It comprises a nearly flat-top core (eq.{[}\ref{eq:UsmallZeta}{]}),
a nearly $1/\zeta$ pedestal (eq.{[}\ref{eq:UintermExp}{]}), and
an exponential foot, eq.(\ref{eq:ULargeZeta}). We already noted that
the escape rate of CR in the foot saturates with the source strength
for $\Pi\gg1$. This is easy to understand as the foot is separated
from the core (source) by the flux controlling pedestal, where the
CR transport is self-regulated in such a way that a fixed CR flux
streams through it to the foot regardless of the strength of the CR
source. The solution is shown in Fig.\ref{fig:SpatialProfileFits}
as a function of $\zeta$ using the self-similar variables $U=\mathcal{P}\sqrt{t}$
and $\zeta=z/\sqrt{t}$. Remarkably, the pedestal portion of the profile
does not change with time also in the physical variables, $\mathcal{P},z$:
$\mathcal{P}\approx2/z$. The core, however, sinks in as $\mathcal{P}\propto1/\sqrt{t}$
but in addition, it expands in $z$ as $\sqrt{D_{{\rm NL}}t}$, to
conserve the integrated pressure contained in the core at the expense
of the pedestal that shrinks accordingly. The latter disappears completely
at $t\sim D_{{\rm ISM}}/D_{{\rm NL}}$, after which the further escape
proceeds in a test particle regime but with a smaller diffusion coefficient
in the core-pedestal region, since waves persist even after most of
particles have escaped. At this point the wave damping should become
important (see Sec.\ref{sec:Comparison-with-other}).

As the test particle approximation is widely used in calculating the
CR escape from their sources, we also show, for comparison, one such
example in Fig.\ref{fig:SpatialProfileFits}. We plot the following
simple test-particle solution 

\begin{equation}
U=\sqrt{\frac{W_{0}}{\pi}}\Pi e^{-W_{0}\zeta^{2}/4}\label{eq:UtestPart}
\end{equation}
that can be obtained from Equation (\ref{eq:Ufit}) or (\ref{eq:Pasympt})
with $V_{0}\left(\Pi\right)$ taken from Equation (\ref{eq:V0cases})
or (\ref{eq:V0Interpol}) for $\Pi\ll1$. To compare the self-regulated
escape with the test-particle one, we formally extend the above solution
to large $\Pi$, as this is normally done within test particle approaches.
Such an extension clearly underestimates the nonlinear level of the
CR pressure by a factor $\sim\Pi^{-1}\exp\left(\Pi/2\right)\gg1$
in the core and in the most of the pedestal. By virtue of lacking
self-regulation, it also overestimates the pressure in the exponential
foot by a factor $\Pi\gg1$.

\subsection{Control parameters and predicted CR flux in physical units \label{sub:EstimatesOfPi}}

The integrated partial pressure $\Pi$ is the most important parameter
that regulates the CR escape. Therefore, we consider it in some detail,
returning to the dimensionful variables and rewriting $\Pi$ as follows
(see Equations {[}\ref{eq:Pnormalization}{]} and {[}\ref{eq:Pidef1}{]}):

\begin{equation}
\Pi\simeq3\frac{C_{{\rm A}}}{c}\frac{a\left(p\right)}{r_{{\rm g}}\left(p\right)}\frac{\bar{P}_{{\rm CR}}\left(p\right)}{B_{0}^{2}/8\pi},\label{eq:PiFinal}
\end{equation}
where $a\left(p\right)$ is the initial size of the CRC, $r_{{\rm g}}$
is the CR gyroradius and $\bar{P}_{{\rm CR}}\left(p\right)$ is their
average partial pressure inside the cloud. Although the ratio of Alfv\'en
speed to speed of light, $C_{{\rm A}}/c$ is typically very small
($\sim10^{-5}-10^{-4}$), the remaining two ratios in the last equation
may be fairly large. Indeed, $a/r_{{\rm g}}$ should amount to several
$c/U_{{\rm ST}}\gg1$, with $U_{{\rm ST}}$ being the shock speed
at the \emph{end }of the Sedov-Taylor stage, as $a$ should exceed
the precursor size $cr_{g}/U_{{\rm ST}}$. Indeed, a significant part
of the shock downstream region filled with CR, that have been convected
there over the entire history of CR production, may significantly
contribute to, if not dominate, the total size of the cloud $a$ \citep{Kang09}.
This is particularly relevant to the CR release during the decreasing
maximum energy (reverse acceleration phase) \citet{Gabici11}. In
the forward acceleration regime (growing maximum energy) Bohm diffusion,
the downstream CR scale height is similar to that of the upstream.
A reasonable estimate for $U_{{\rm {\rm ST}}}$ is $c/U_{{\rm ST}}\gtrsim10^{3}$.
Finally, the CR pressure in the source should considerably exceed
the magnetic pressure as both quantities are roughly in equipartition
in the background ISM. Alternatively, as is usually assumed, the accelerated
electrons are in equipartition with the magnetic energy inside the
source, since they should have lost their excessive energy via synchrotron
radiation. As electrons are thought to be involved in the acceleration
at $\sim10^{-2}$ of the proton level, the last ratio in Equation
(\ref{eq:PiFinal}) is then $\sim10^{2}$, thus giving, perhaps, an
upper bound to the pressure parameter $\Pi\sim10^{2}/M_{A}$, where
$M_{A}=U_{{\rm ST}}/C_{A}$.

One of a few other ways the parameter $\Pi$ can be looked at is to
express it through the average acceleration efficiency $\mathcal{E}=2\bar{P}_{{\rm CR}}/\rho\bar{U}^{2}$,
where $\bar{U}$ is the shock velocity, appropriately averaged over
the CR acceleration history (one may expect $\bar{U}\gg U_{{\rm {\rm ST}}})$.
Then $\Pi\simeq3\left(\bar{U}^{2}/cC_{{\rm A}}\right)\left(a/r_{{\rm g}}\right)\mathcal{E}\simeq3A\left(\bar{U}^{2}/U_{{\rm {\rm ST}}}C_{{\rm A}}\right)\mathcal{E}$,
where we have introduced a factor $A=\left(a/r_{{\rm g}}\right)U_{{\rm ST}}/c\gtrsim1$.
In this form, the estimate indeed boils down to the acceleration efficiency
$\mathcal{E}$ with the remaining quantities ($A,\bar{U}$ and $U_{{\rm ST}}$)
being more accessible to specific models. Since the acceleration efficiency
$\mathcal{E}$ is believed to be at least $\gtrsim0.1$ for productive
SNRs, we conclude that the control parameter $\Pi$ is rather large
than small. The caveat here is that it might, in some cases, be too
large to limit the applicability of the above treatment. We return
to this issue in the Discussion section. 

Once the major control parameter is known, we can calculate the distribution
function $f_{{\rm CR}}$ of the CR, released into the ISM, depending
on the distance from the source $z$ and time, using the parametrization
in Equation (\ref{eq:Ufit}). It is convenient to rewrite it in the
form of the CR partial pressure $\hat{P}_{{\rm CR}}$ in physical
units as follows

\begin{equation}
\frac{\hat{P}_{{\rm CR}}}{B_{0}^{2}/8\pi}\simeq0.8\left(\frac{M_{{\rm A}}}{A}\frac{10^{4}{\rm yrs}}{t}\frac{n}{{\rm 1cm^{-3}}}\frac{E}{{\rm 1GeV}}\right)^{1/2}\left(\frac{1\mu{\rm G}}{B}\right)^{3/2}U\left(\zeta\right)\label{eq:PcrFin}
\end{equation}
Here $M_{{\rm A}}=U_{{\rm ST}}/C_{{\rm A}}$, $n$ is the plasma number
density, $E$ is the particle energy and $B$ is the magnetic field.
The self-similar coordinate $\zeta$ can be represented, in turn,
in the following way

\begin{equation}
\zeta\simeq17\left(\frac{M_{{\rm A}}}{A}\frac{{\rm 1GeV}}{E}\frac{B}{1\mu{\rm G}}\frac{10^{4}{\rm yrs}}{t}\right)^{1/2}\frac{z}{{\rm pc}}\label{eq:zetaFin}
\end{equation}
Once this quantity is calculated for given $z$ and $t$, the CR partial
pressure in Equation (\ref{eq:PcrFin}) can be obtained from Equation
(\ref{eq:Ufit}) or from Fig.\ref{fig:SpatialProfileFits}. Note that
the scale of the CR pressure is determined by the maximum of $U=U\left(0\right)=\sqrt{2W_{0}}V_{0}\exp\left(\Pi/2\right)$,
where $V_{0}$ is given by Equation (\ref{eq:V0Interpol}). The CR
partial pressure also depends on the energy through $W_{0}$ and $\Pi$,
apart from the factor $\sqrt{E}$ in Equation (\ref{eq:PcrFin}),
which will be discussed below. Finally, the background diffusion coefficient
of CR in physical units is related to the dimensionless diffusivity
$W_{0}^{-1}$ as follows

\[
D_{{\rm ISM}}=6.6\cdot10^{27}\frac{10^{-4}}{W_{0}}\frac{a}{{\rm 1pc}}\frac{B}{1\mu G}\left({\rm \frac{1cm^{-3}}{n}}\right)^{1/2}\frac{{\rm cm^{2}}}{{\rm sec}}.
\]

\subsection{The Form of the Spectrum of escaping CR\label{sub:Spectrum-of-escaping}}

As the transition from the flat-top part of the normalized (Equation
{[}\ref{eq:Pnormalization}{]}) partial pressure $\mathcal{P}\left(p\right)$
to the pedestal is described by a function $\mathcal{P}\approx2$$\left\{ z^{5/3}+\left[D_{{\rm NL}}\left(p\right)t\right]^{5/6}\right\} ^{-3/5}$,
the pressure $\mathcal{P}\left(p\right)$ is almost $p$-independent
(the corresponding particle momentum distribution $f_{{\rm CR}}\propto\kappa_{{\rm B}}p^{-4}/\hat{z}$,
in the physical units, $\hat{z}=az$) for such momenta where $D_{{\rm NL}}\left(p\right)<z^{2}/t$,
for the fixed dimensionless observation point $z$ and time $t$.
At $p=p_{{\rm br}}$, where $D_{{\rm NL}}\left(p_{{\rm br}}\right)=z^{2}/t$,
the spectrum incurs a break. If we approximate $D_{{\rm NL}}\propto p^{\delta}$
near $p\sim p_{{\rm br}}$, the break will have an index $\delta/2$,
as may be seen from the above formula for $\mathcal{P}$. The index
$\delta$ thus derives from the momentum dependence of the $D_{{\rm ISM}}\left(p\right)$
and from that of the coordinate integrated CR pressure $\Pi\left(p\right)$
(through the factor $\exp\left(-\Pi\right)$ in equation {[}\ref{eq:DNLdef}{]}).
Furthermore, if we represent $\exp\left(-\Pi\right)\propto p^{-\sigma}$
and $D_{{\rm ISM}}\propto p^{\lambda}$ at $p\sim p_{{\rm br}}$,
so that $\delta=\lambda-\sigma$, then $\mathcal{P}$ is flat at $p<p_{{\rm br}}$
for $\delta>0$ and steepens to $p^{-\delta/2}$ at $p=p_{{\rm br}}$.
Conversely, if $\delta<0$, $\mathcal{P}$ raises with $p$ as $p^{-\delta/2}$
at $p<p_{{\rm br}}$ and it levels off at $p>p_{{\rm br}}$. Note,
however, that $\Pi$ is momentum independent in the important case
of the $p^{-4}$ distribution of the initial CRC with the scale height
$a\left(p\right)\propto\kappa_{{\rm B}}$. This case corresponds to
the test-particle DSA spectrum, $\propto p^{-4}$ with a scale height
$a$ estimated from the shock precursor size. The break has then an
index $\lambda/2$ and it is entirely due to the $D_{{\rm ISM}}$
momentum dependence.

\section{Comparison with other approaches\label{sec:Comparison-with-other}}

Given the variety of approaches to the CR escape, we extend our brief
discussion in the Introduction section by putting our treatment into
perspective. Most of the approaches can be categorized into the following
three kinds. First of all, a simple test particle approach (TP) is
feasible for a rarefied CRC, when wave generation is negligible \citep[e.g., ][]{AharAt96,DruryEscape11,OhiraProp11}.
It solves the \emph{linear }diffusion equation for the CRs with a
diffusion coefficient determined by a given (e.g., background) turbulence.
Next, there are modifications of this approach that include the wave
generation by escaping particles balanced by the MHD cascades \citep[e.g., ][]{PtusZir05,PtuskinNLDIFF08,YanLazarianEscape12}.
The wave intensity, thus obtained, is then used to calculate the particle
diffusion coefficient and their distribution. We call these approaches
modified TP, as they do not evolve waves but slave them to the instant
particle distribution.The third treatment is based on a quasi-linear
(QL) mechanism, whereby the wave stabilization primarily occurs through
their backreaction on the particles (e.g., suppression of instability
by pitch-angle isotropization). The nonlinear effects may or may not
be significant during the QL stabilization. As the QL wave saturation
operates at the lowest-order in wave energy, it is imperative to consider
it first, particularly where the waves become weaker. This approach
has been well tested on a somewhat similar problem of a hot electron
cloud expansion into a plasma \citep{RyutovSagdeev70,IvanovSagdeev70}.
In the case of a CRC expansion into the ISM, there is an extended
region (i.e. ``pedestal'' self-confinement region) where the level
of turbulence significantly exceedsthe backgroundbut is still not
high enough for the nonlinear effects to dominate over the quasilinear
ones. As our results show, this is the most important domain that
determines the spectrum of escaping CRs. Simply put, the dynamics
is dominated by wave generation and self-regulated particle escape,
rendering the nonlinear wave dynamics and MHD cascades less important.
It should be noted, however, that the instability driving pitch-angle
anisotropy is supported by the spatial inhomogeneity of the CRC, so
that the full stabilization occurs (under negligible damping) only
when particles spread to infinity. 

It is nonetheless worthwhile to consider the possible effect of wave
damping we have neglected. For example, \citet{PtuskinNLDIFF08},
while neglecting $dI/dt$ on the l.h.s of eq.(\ref{eq:dIdt}) balance
the driving term with the damping term on its r.h.s and assume a Kolmogorov
dissipation for $\Gamma$, 
\begin{equation}
\Gamma=kC_{A}\sqrt{I}/\left(2C_{K}\right)^{3/2}\label{eq:GammaPZ}
\end{equation}
with $C_{K}\approx3.6$ and $k\simeq1/r_{g}\left(p\right)$ being
the resonant wave number. Therefore, only one equation (\ref{eq:dPdt})
needs to be solved. The CR density decays then at the source as $\propto t^{-3/2}$
and the flat-topped, self-confined part of the CR distribution spreads
as $z\propto t^{3/2}$, both pointing at the superdiffusive CR transport%
\footnote{Recall that our results give for these quantities $t^{-1/2}$ and
$t^{1/2}$, respectively%
}. The reason is clearly in a very strong wave damping due to the Kolmogorov
dissipation. For the same reason this solution does not recover the
test particle asymptotic $P_{CR}\propto t^{-1/2}\exp\left(-z^{2}/4D_{{\rm ISM}}t\right)$,
physically expected in $z,t\to\infty$ limit in the interstellar medium
with the background diffusion coefficient $D_{{\rm ISM}}$. While
such a strong wave damping may be justified in the core of the CRC
during the early phase of escape, the overall confinement is controlled
by the $1/z$-pedestal, where the waves are relatively weak and the
Kolmogorov cascade can hardly be important. Moreover, since the pedestal
plays a role of a barrier enclosing the core, the wave-particle interaction
dynamics in the core is less important than that in the pedestal.
In this regard, the particle distribution in the core is similar to
the QL plateau on particle distribution in momentum space. When established,
the plateau does not depend on the interaction of the waves that create
it.

An alternative choice of damping mechanism is the \citet{goldr97}
(GS) MHD cascade, which seems to be more appropriate in $I\lesssim1$
regime \citep{FarmerGoldr04,BeresnLaz08,YanLazarianEscape12} and
may play some role in the pedestal. The damping rate is 
\begin{equation}
\Gamma=C_{A}\sqrt{\frac{k}{L}}\label{eq:GammaGS}
\end{equation}
where $L$ is the outer scale of turbulence which may be as large
as $100pc$. Not only is this damping orders of magnitude (roughly
a factor $\sqrt{L/r_{g}}$ ) lower than the Kolmogorov one but, it
does not depend on $I$ and can be considered as coordinate independent.
Then, the damping term does not violate the quasilinear integral,
eq.(\ref{eq:QLint}). The dimensionless equation (\ref{eq:dwdt})
can thus be rewritten simply as follows (outside the initial CRC)
\[
\frac{\partial W}{\partial t}=\frac{\partial}{\partial z}\frac{1}{W}\frac{\partial W}{\partial z}-\Gamma^{\prime}W,
\]
and the dimensionless damping $\Gamma^{\prime}=a/\sqrt{r_{g}L}\sim\left(c/U_{{\rm ST}}\right)\sqrt{r_{g}/L}$
may be eliminated by replacing $W\exp\left(\Gamma^{\prime}t\right)\to W$,
$\intop_{0}^{t}\exp\left(\Gamma^{\prime}t\right)dt\to t$. Our results
then remain intact in the relabelled $t$ and $W.$

\begin{figure}
\includegraphics[scale=0.6]{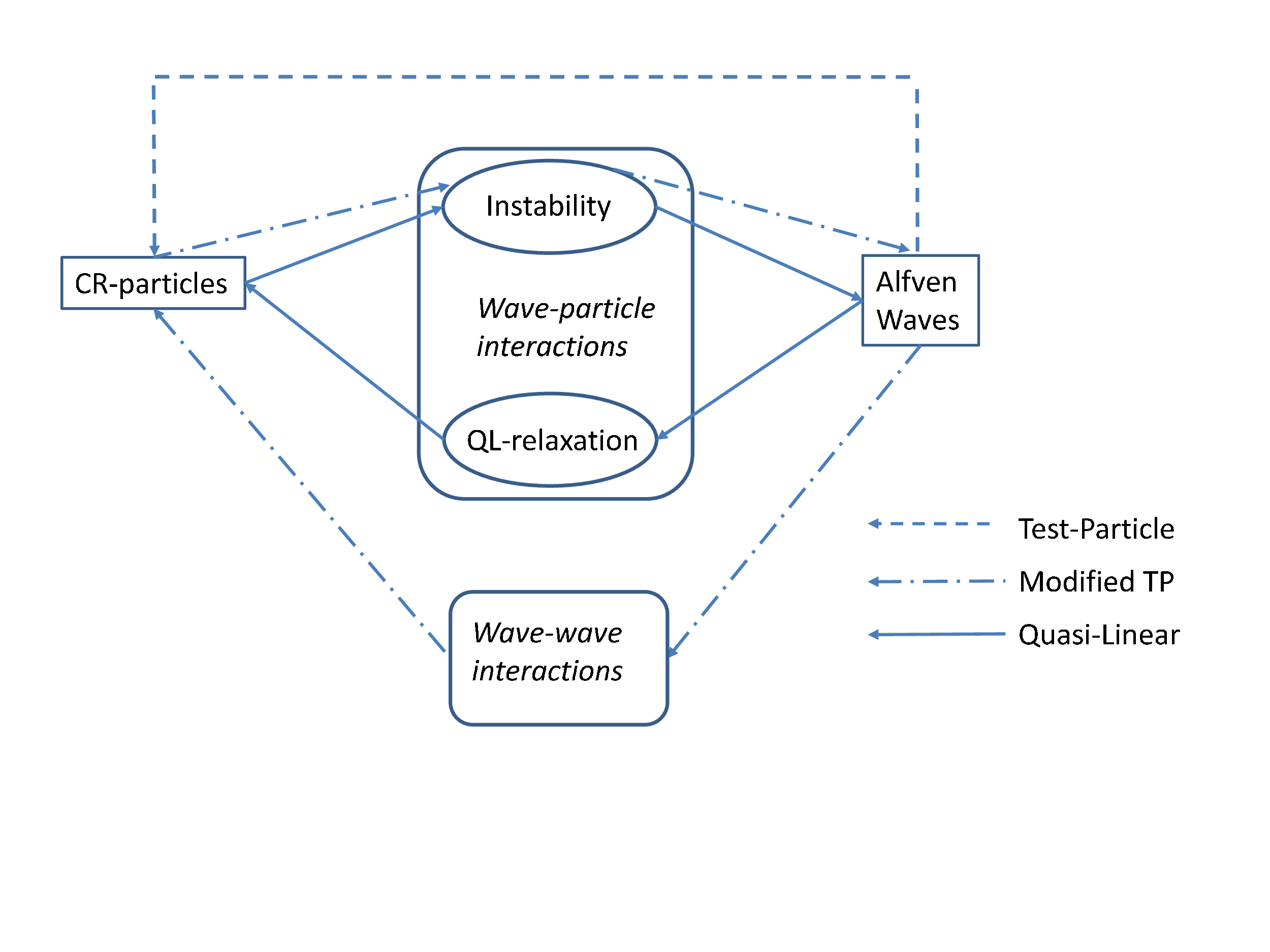}

\caption{Three different approximations for studying injection of accelerated
CRs into ISM: Test particle (TP), modified TP (with unstable wave
growth and nonlinear wave evolution), and quasi-linear (QL) (with
self-consistent time dependent wave-particle interactions). \label{fig:Chart}}
\end{figure}

The above three approaches to the CR escape are summarized in Fig.\ref{fig:Chart}.
The key ingredients are the CR particles and Alfv\'en waves while the
relevant physical phenomena are the wave-particle and wave-wave interactions.
Under the latter one may loosely understand both weakly-turbulent
parametric processes (such as wave decay) and turbulent cascades similar
to those described above. The wave-particle interactions comprise
the unstable growth of Alfv\'en waves together with their back-reaction
on the CRs. In principle, the wave-wave interactions need to be included
in the quasi-linear treatment, particularly in the core of the CRC
when $w_{{\rm max}}\gtrsim1$. However, as the wave dynamics in the
flat core is relatively unimportant, one may employ the Bohm diffusion
coefficient in this region, i.e. where $z^{2}/t\lesssim D_{{\rm {\rm NL}}}$
by setting $D_{{\rm NL}}\sim\kappa_{{\rm B}}$ and thus obtain a result
already given by Equation (\ref{eq:UBohm}). Formally, it is equivalent
to the test-particle approximation with the minimum diffusion coefficient
set at the Bohm level. 

To summarize our results, we have considered the self-consistent relaxation
of a CR cloud (CRC) injected into a magnetized plasma (ISM) under
the assumption of an initially weak background turbulence, $\left(\delta B/B_{0}\right)^{2}\ll1$,
so that the cross-field diffusion is negligible, $\kappa_{\perp}\ll\kappa_{\parallel}$
\emph{outside }the cloud and the particles escape largely along the
field, i.e., in $z$-direction. The principal parameter that regulates
the CR escape from the cloud is identified to be the coordinate-integrated
partial pressure $\Pi$, given \eg in Equation (\ref{eq:PiFinal}).
Resonant waves of the length $\sim r_{{\rm g}}^{-1}\left(p\right)$,
driven by the run-away cloud particles, are found to confine the core
CRs very efficiently when $\Pi\gg1$. 

The resulting normalized (Equation {[}\ref{eq:Pnormalization}{]})
CR partial pressure profile $\mathcal{P}$ comprises the following
three zones: (i) a quasi-plateau (core) at $z/\sqrt{t}<\sqrt{D_{{\rm NL}}}$
of a height $\sim\left(D_{{\rm NL}}t\right)^{-1/2}$, which is elevated
by a factor $\sim\Pi^{-1}\exp\left(\Pi/2\right)\gg1$, compared to
the test particle solution because of the strong quasi-linear suppression
of the CR diffusion coefficient with respect to its background (test
particle) value $D_{{\rm ISM}}$: $D_{{\rm NL}}\sim D_{{\rm ISM}}\exp\left(-\Pi\right)$
(ii) next to the core, where $\sqrt{D_{{\rm NL}}}<z/\sqrt{t}<\sqrt{D_{{\rm ISM}}}$,
the profile is scale invariant, $\mathcal{P}\approx2/z$. The CR distribution
in this ``pedestal'' region is fully self-regulated and independent
of $\Pi$ and $D_{{\rm ISM}}$ for $\Pi\gg1$, (iii) the tail of the
distribution at $z/\sqrt{t}>\sqrt{D_{{\rm ISM}}}$ is similar in shape
to the test particle solution but it saturates with $\Pi\gg1$, so
that the CR partial pressure is $\propto\left(D_{{\rm ISM}}t\right)^{-1/2}\exp\left(-z^{2}/4D_{{\rm ISM}}t\right)$,
independent of the strength of the CR source $\Pi$, in contrast to
the test-particle result that scales as $\propto\Pi$. 

Depending on the functions $\Pi\left(p\right)$ and $D_{{\rm ISM}}\left(p\right)$,
the resulting CR spectrum generally develops a spectral break for
the fixed values of $z$ and $t$ such that $z^{2}/t\sim D_{{\rm NL}}\left(p\right)\sim D_{{\rm ISM}}\exp\left(-\Pi\right)$.

\section{Discussion and Outlook}

The CR escape from both active and fading accelerators (old SNR) is
being actively studied through direct observations of CR illuminated
molecular clouds \citep{AharW28HESS08,AbdoW28_10,AgileW44_11,MagicW51_12,UchiW44_12,FermiSciW4413}.
To date, most of the information is obtained from the old remnants
and they consistently show a broad spectrum of CR escape. This is
clearly at odds with an intuitive high energy biased escape, seemingly
justified by the higher mobility of energetic CRs. Indeed, as CR diffusion
coefficient grows with momentum, the test-particle solution predicts
a low-energy cutoff to be present due to the factor $\exp\left[-z^{2}/4D_{{\rm ISM}}\left(p\right)t\right]$
in the CR distribution at a certain distance $z$ from the source.
In a combination with a steep power-law or a favorably placed upper
cutoff, the escape flux narrowly accumulates towards the maximum energy.
The momentum dependent CR mobility underlies most of the current CR
escape models \citep[e.g., ][]{PtusZir05,ZirakPtusEsc08,GabiciAharEsc09}. 

Although the same exponential factor is present in the QL solution
obtained in this paper, it pertains only to the farthermost zone,
where the CR partial pressure is much lower (by factor $\Pi^{-1}\ll1$)
than the test particle prediction for the same distance from the source
and time elapsed from the CR release. Closer to the accelerator, in
an extended scale-invariant zone where the CR level is much higher
and decays as slowly as $1/z$, the escape mechanism is different
from the one controlled by the mere energy dependence of the CR diffusivity.
It is self-regulated in such a way that, if particles leak excessively
in some energy range, they also drive stronger resonant waves to reduce
their leakage and vice versa. As a result, the overall escape spectrum
relaxes roughly to an equipartition of the CR partial pressure in
momentum, e.g., $f_{{\rm CR}}\propto p^{-4}$ (with important deviations
described in Sec.\ref{sub:Spectrum-of-escaping}), which also balances
the driver (gradient of CR partial pressure) with the generated waves.
No low-energy leakage suppression therefore occurs. The fundamental
difference of this leakage mechanism from the test particle one is
that it is entirely controlled by self-generated rather than prescribed
waves, or by waves derived from other energy sources, such as ambient
MHD cascades. That is why it predicts energetically much broader leakage
than many other approaches do \citep[e.g., ][]{PtusZir05,EllisonBykEscape11}

It should be admitted that the self-confinement solution obtained
in this paper is strictly valid for a stopped accelerator, so that
there is no CR energy growth and strong plasma flows, such as those
found near shocks. Therefore, care should be exercised in comparing
this solution with the standard DSA predictions. At the same time,
even if CRs were escaping from the DSA through a free-escape boundary
(FEB) or an upper momentum cutoff,they would propagate further out
diffusively. Yet their escape is often enforced by imposing an ad
hoc sudden jump in the diffusion coefficient $D\left(p,z\right)$
at a specific momentum $p$ (or FEB position $z$). Despite this jump,
the CR phase space density must be continuous,%
\footnote{This statement is strictly valid for a FEB imposed by enhanced diffusion
in coordinate space. In the case of a jump in momentum \citep[e.g.,][]{PtusZir05},
the situation is more complicated in that the particle distribution
should become increasingly anisotropic in pitch angle, as the escape
is assumed one-sided (upstream). However, an introduction of weak
Fermi-II acceleration (diffusion in momentum) would validate this
statement also in this case.%
} and for strong CR sources still high enough to drive waves while
the CR escape. From this point on, the solution obtained in this paper
may be applied and compared with the DSA predictions, as the waves
are driven locally both in momentum and in coordinate space. The only
relevant requirement is that particles do not interact with the shock,
as implied in most escape models in the first place. But, the feedback
from the self-generated waves on the CR escape is not included in
the models that predict a peaked energy escape.

Furthermore, the self-regulated escape solution shows a gradual increase
of the CR diffusion from a low (Bohm) to high (TP) regime, across
the region where $P_{CR}\propto1/z$ and the CR diffusion coefficient
$D_{{\rm CR}}\propto z^{2}$, thus keeping the flux$-D_{{\rm CR}}\nabla\mathcal{P}\approx{\rm const}$.
Both scalings are clearly inconsistent with a sudden jump in $D_{{\rm CR}}$
with the corresponding jump in $\nabla P_{CR}$. Moreover, collapsing
the scale-invariant region to a point (FEB) would not only change
the CR escape flux considerably but, also an extended region of enhanced
CR pressure would have been lost (see Fig.\ref{fig:SpatialProfileFits}).
This region (which we loosely dubbed ``pedestal'' by analogy with
the improved confinement regimes in magnetic fusion devices, primarily
in tokamaks\citep[e.g.][]{Wagner82,HintonStab93,DiamLeb95,MDtranspBif08,Maggi10})
may be detectable when it overlaps with molecular clouds. According
to the test-particle theory the CR density decays as $t^{-d/2}$ in
the region of their initial release, with $d$ being the dimensionality
of the escape ($d=1$ for escape along the field). In the self-regulated
escape, the CR density stays constant in time in the pedestal region,
where $P_{{\rm CR}}\propto1/z$, until this region is overwhelmed
by the expanding central plateau where $P_{{\rm CR}}$ decays as $1/\sqrt{t}$.
Before it happens, the CR density is higher than in the test-particle
case, Fig.\ref{fig:SpatialProfileFits}, which is a prediction that
may soon become testable. 

Recent detailed observations of the SNR W44, W51C, IC 443 and W28,
surrounded by MC, provide good examples to study possible CR escape
scenarios \citep{AharW28HESS08,Abdo10W44full,AbdoIC443_10,AbdoW28_10,AgileW44_11,MagicW51_12,UchiW44_12}.
First of all, they almost invariably show spectral breaks that, however,
may be understood in terms of the interaction of accelerated protons
with a partially ionized dense gas \citep{MDS05,MDS_11NatCo}. The
indices below and above the breaks are consistent with the following
two scenarios. Namely, CR protons may reach the molecular cloud while
still being accelerated at a SNR shock, or they may escape with a
similar spectrum, $p^{-4}$. As we have shown in the present paper,
such escape occurs via the CR interaction with self-generated waves.
Another important aspect of these observations regards the morphology
of the interaction. Of particular interest is the recent analysis
of Fermi-LAT results revealing two bright spots of gamma emission
adjacent to the central source in W28 \citep{UchiW44_12}. Their distinct
bi-polar appearance may be indicative of a CR escape along the local
magnetic field.

Note that the anisotropic diffusion of cosmic rays in the form of
bipolar CRC may result in quite specific morphologies of extended
gamma-ray images - the imprints of cosmic rays interacting with the
surrounding diffuse gas are generally concentrated in dense molecular
clouds. Since the gamma-ray flux is proportional to the product of
densities of CRs and the diffuse gas, we should expect a rather general
correlation between the gamma-ray fluxes and the column densities
of the interstellar gas. However, it is obvious that in the case of
propagation of bipolar CR clouds through an inhomogeneous clumpy gaseous
environment, the gamma-ray intensity contours can significantly deviate
from the CO and 21cm maps characterizing the spatial distributions
of the molecular and atomic gases, respectively. At the same time,
the brightest parts of the spatial distribution of gamma-rays should
correspond to the regions where the CR cloud overlaps with dense gas
clouds \emph{inside the magnetic flux tube} connected with the CR
source.

\acknowledgements{}

Support by the Department of Energy, Grant No. DE-FG02-04ER54738 is
gratefully acknowledged. P.H.D. acknowledges support from WCI Program
of the National Research Foundation of Korea (NRF) (NRF Grant Number:
WCI 2009-001).

\appendix{}

\section{Details of self-similar solution\label{sec:Details-of-self-similar}}

The full expressions for the solution $w\left(\zeta\right)$ represented
in a short form by eqs.(\ref{eq:wOfV}) and $(\ref{eq:zetaOfV})$
can be written as follows

\begin{equation}
w=W_{0}\left\{ \begin{array}{cc}
e^{\int_{0}^{V}dV/\sqrt{R\left(V\right)}}, & \zeta\ge\zeta_{1}\\
e^{\int_{0}^{V_{1}}dV/\sqrt{R\left(V\right)}+\int_{V}^{V_{1}}dV/\sqrt{R\left(V\right)}}, & 0<\zeta<\zeta_{1}
\end{array}\right.\label{eq:wofVfull}
\end{equation}
Throughout this Appendix, the positive branch of $\sqrt{R}$ is used.
In some cases, it is convenient to represent $w$ in the domain $0<\zeta<\zeta_{1}$
as 

\begin{equation}
w=w_{{\rm max}}e^{-\int_{V_{0}}^{V}dV/\sqrt{R\left(V\right)}}\label{eq:Apwmax}
\end{equation}
where $V_{0}\equiv\exp\left(-q/2\right)$, \ie $\zeta\left(V_{0}\right)=0$,
and

\begin{equation}
w_{{\rm max}}=W_{0}e^{\int_{0}^{V_{1}}dV/\sqrt{R\left(V\right)}+\int_{V_{0}}^{V_{1}}dV/\sqrt{R\left(V\right)}}.\label{eq:wmaxAp}
\end{equation}
For the coordinate $\zeta$ we have, Fig.\ref{fig:FunctionRofV}:

\begin{equation}
\zeta=\sqrt{\frac{2}{w}}\left\{ \begin{array}{cc}
V+2\sqrt{R\left(V\right)} & \zeta\ge\zeta_{1}\\
V-2\sqrt{R\left(V\right)} & 0\le\zeta<\zeta_{1}
\end{array}\right.\label{eq:zetaAp}
\end{equation}
The function $w\left(\zeta\right)$, implicitely defined by eqs.(\ref{eq:wofVfull})
and (\ref{eq:zetaAp}) is shown in Fig.\ref{fig:Analytic-vs-numerical}(see
Sec.\ref{sub:SpatialProfileOfP} for approximate explicit results).
The partial CR pressure can be represented according to eq.(\ref{eq:PofW})
as follows

\[
\mathcal{P}\left(z,t\right)=\left.\sqrt{2w\left(\zeta\right)/t}V\left(\zeta\right)\right|_{\zeta=z/\sqrt{t}},\;\;\; z>1
\]
 where $V\left(\zeta\right)$ is defined by eq.(\ref{eq:zetaAp}).

\section{Half-life of the CR cloud\label{sec:Half-life-of-CR}}

Equation (\ref{eq:HalfLifetDef}) can be continued as

\begin{equation}
\Pi_{1}\left(t_{1/2}\right)=\frac{1}{2}\Pi=\frac{1}{2}\ln\frac{w_{{\rm max}}}{W_{0}}\equiv\ln\frac{w_{1/2}}{W_{0}}\label{eq:PiAp}
\end{equation}
where we have introduced a ``half-life'' amplitude $w_{1/2}$, which
can also be associated with a point $V_{1/2}$ (see Equation{[}\ref{eq:wOfV}{]})

\[
\ln\frac{w_{1/2}}{W_{0}}=\int_{0}^{V_{1/2}}dV/\sqrt{R\left(V\right)},
\]
and with the corresponding self-similar coordinate $\zeta_{1/2}=1/\sqrt{t_{1/2}}$.
In other words, $w_{1/2}=w\left(\zeta_{1/2}\right)=w\left(V_{1/2}\right)$.
Note that $w_{1/2}$ can be represented as a geometric mean $w_{1/2}\equiv\sqrt{W_{0}w_{{\rm max}}}$
(cf. equation {[}\ref{eq:Pidef2}{]}), so that in terms of the function
$\ln w\left(\zeta\right)$, the point $\zeta_{1/2}$ corresponds to
the FWHM. The quantity $V_{1/2}$ can thus be obtained from the following
equation

\begin{equation}
2\int_{0}^{V_{1/2}}dV/\sqrt{R\left(V\right)}=\int_{0}^{V_{1}}dV/\sqrt{R\left(V\right)}+\int_{V_{0}}^{V_{1}}dV/\sqrt{R\left(V\right)}\label{eq:VhalfDef}
\end{equation}
In the case of a weak CRC, $\Pi\ll1$, we obviously have $V_{1/2}<V_{0}\sim V_{1}\ll1$.
Calculating the integrals in this limit and substituting $R$ from
equation (\ref{eq:FirstInt}) we have

\[
{\rm erfc}\left(\ln\frac{1}{V_{1/2}}-\frac{q}{2}\right)^{1/2}=\frac{1}{2}
\]
or

\[
V_{1/2}=e^{-\frac{q}{2}-\sigma}=V_{0}e^{-\sigma}\approx V_{1}e^{-\sigma}/\left(1+\frac{1}{2}e^{-q}\right)
\]
where ${\rm erfc}\left(\sigma^{1/2}\right)=1/2$, so that $\sigma\approx0.23$.
Substituting then $V_{1/2}$ into $\zeta_{1/2}$ from equation (\ref{eq:zetaAp})
we obtain the result given by equation (\ref{eq:thalflin}).

In the case $\Pi\gg1$, the integrals on the r.h.s. of Equation (\ref{eq:VhalfDef})
are dominated by the upper limit, so the integral on the l.h.s. must
also be and we deduce $V_{1/2}\approx V_{1}$, from which we obtain
the nonlinear CR half-life in Equation (\ref{eq:thalfLINNL}).

\clearpage

\end{document}